\documentclass[french,english,authoryear, 12, preprint]{elsarticle}
\usepackage[T1]{fontenc}
\usepackage[latin9]{inputenc}
\usepackage{geometry}
\usepackage{xcolor}
\usepackage{pdfcolmk}
\usepackage{array}
\usepackage{textcomp}
\usepackage{amstext}
\usepackage{graphicx}
\usepackage{setspace}
\PassOptionsToPackage{normalem}{ulem}
\usepackage{ulem}
\makeatletter

\newcommand*\LyXZeroWidthSpace{\hspace{0pt}}
\providecommand{\tabularnewline}{\\}
\providecolor{lyxadded}{rgb}{0,0,1}
\providecolor{lyxdeleted}{rgb}{1,0,0}

\usepackage{hyperref}

\@ifundefined{showcaptionsetup}{}{%
 \PassOptionsToPackage{caption=false}{subfig}}
\usepackage{subfig}
\makeatother

\usepackage{babel}
\makeatletter
\addto\extrasfrench{%
   \providecommand{\fg}{\ifdim\lastskip>\z@\unskip\fi~\frqq}%
}

\makeatother
\begin{document}

\begin{frontmatter}{}

\title{Ice state evolution during spring in Richardson crater, Mars }

\author{Fran\c{c}ois Andrieu$^{1,2}$, Fr{\'e}d{\'e}ric Schmidt$^{1}$,
Sylvain Dout{\'e}$^{3}$, Eric Chassefi{\`e}re$^{1}$}

\address{$^{1}$ GEOPS, Univ. Paris-Sud, CNRS, Universit{\'e} Paris-Saclay,
Rue du Belv{\'e}d{\`e}re, B{\^a}t. 504-509, 91405 Orsay, France,
$^{2}$LESIA, Observatoire de Paris, PSL Research University, CNRS,
Sorbonne Universités, UPMC Univ. Paris 6, Univ. Paris Diderot, Sorbonne
Paris Cité, France, $^{3}$IPAG, Univ. Grenoble Alpes, CNRS, F-38000
Grenoble, France}
\begin{abstract}
The Martian climate is governed by an annual cycle, that results in
the condensation of CO$_{2}$ ice during winter, up to a meter thick
at the pole and thousands of kilometers in extension. Water and dust
may be trapped during the condensation and freed during the sublimation.
In addition, ice may be translucent or granular depending on the deposition
process (snow vs direct condensation), annealing efficiency, and dust
sinking process. The determination of ice translucency is of particular
interest to confirm or reject the cold jet model (also known as Kieffer
model).

This work is focused on the dune field of Richardson Crater (72\textdegree S,
180\textdegree W)in which strong interactions between the water, dust
and CO$_{2}$ cycles are observed. We analyzed CRISM hyperspectral
images in the near IR using radiative transfer model inversion. We
demonstrate that among the states of CO$_{2}$ ice, the translucent
state is observed most frequently. The monitoring of surface characteristics
shows a decrease in the thickness of the ice during the spring consistently
with climate models simulations. We estimate a very low dust content
of a few ppmv into the CO$_{2}$ ice, consistent with the formation
scenario of cold jets. The water impurities is around 0.1\%v, almost
stable during the spring, suggesting a water escape from the surface
of subliming CO$_{2}$ ice layer. The water ice grain size varies
in a range 1 to 50 microns. From these results, we propose the following
new mechanism of small water ice grain suspension: as a cold jet occurs,
water ice grains of various sizes are lifted from the surface. These
jets happen during daytime, when the general upward gas flux from
the subliming CO$_{2}$ ice layer is strong enough to carry the smaller
grains, while the bigger fall back on the CO$_{2}$ ice layer. The
smaller water grains are carried away and integrated to the general
atmospheric circulation. 
\end{abstract}

\end{frontmatter}{}

\section{Introduction}

The Martian climate is mostly governed by a planetary scale CO$_{2}$
condensation-sublimation cycle. As the atmosphere is composed of 96\%
CO$_{2}$ \citep{Mahaffy2013,Owen1977}, the CO$_{2}$ cycle is the
dominant climatic cycle \citep{Leighton1966,Tillman1988,Mischna2003}.
The other two cycles that play important role in the martian climate
are the dust cycle and the water cycle. General circulation models
for Mars are now able to predict at a regional scale the amount of
CO$_{2}$ ice present at the surface, but the water ice and dust cycles
are still under-constrained. On Mars, climatic quantities such as
CO$_{2}$ ice quantity at the surface are computed into boxes that
are 3.75\textdegree{} in latitude and 5.25\textdegree{} in longitude
(up to a few hundreds of kilometers). As approximately 30\% of the
atmosphere condenses at the surface during winter, the CO$_{2}$ cycle
has a huge radiative impact on the planet and it is therefore of prime
importance to be able to model it correctly. 

One method to decipher the microphysics of the surface/atmosphere
exchange involved in this cycle is to study the details of the reflectance
response of the soil. Indeed, the full shape of the absorption bands
contains informations about the surface and atmospheric properties.
Not only the chemical composition (presence/absence of CO$_{2}$ or
water ice) can be derived, but also the surface properties (grains
size, abundances...). Various models have been proposed to simulate
the reflectance of the Martian surface when covered with CO$_{2}$
ice \citep{Doute1998,Appere2011,singh2016improved}, but a common
point of those models is that they all considered that the ice deposits
are granular, with a porosity close to 50\%. Yet, defrosting spring
activity \citep{Piqueux2003,Kieffer2006,Hansen,andrieu2014co2} strongly
suggest that during most of the spring (defrosting season), the CO$_{2}$
ice layer is in fact in a translucent slab state. Then it is very
important to be able to model the radiative response of such a surface,
very different from a granular surface's response.

Recently, we have built a radiative transfer model able to simulate
the near infrared reflectance of a surface constituted of a slab of
CO$_{2}$ ice, clean or contaminated with water ice and mineral dust,
and that lays on the Martian regolith \citep{Andrieu2015}. We also
developed an inversion method, adapted to our study that we validated
in a previous study: based on Bayesian formalism, the method uses
a reference theoretical database to estimate a probability density
function of the value of the parameters that are to be investigated\citep{Andrieu2016}.
The combination of the inversion method with the radiative transfer
model allows to retrieve ice covered surface properties from their
near infrared spectrum. The properties retrieved are the thickness
of the ice layer, and the proportions and grain sizes of impurities.
These tools allow us to follow the volatiles exchanges between the
Martian surface and atmosphere during spring, to bring better constraint
to the seasonal Martian climatic cycles.

This study focuses on the dune field inside Richardson Crater (180\textdegree W,
72\textdegree S). This dune field was widely studied \citep{Mohlmann2010,Kereszturi2011,Supulver2001,Edgett2000,MARTINEZ2012}
because it showed intense spring activity and interactions between
water and CO$_{2}$ cycle. 

In this work we:
\begin{itemize}
\item argue about the presence/absence of translucent CO$_{2}$ ice.
\item determine the surface properties of the seasonal frost (thickness,
grain-size, dust impurities, water impurities)
\item propose a microphysical model to describe the exchanges of volatiles
in Richardson crater
\end{itemize}

\section{Method}

In this work we compare real data with synthetic data to perform an
``inversion''. To produce realistic synthetic data, we first need
a realistic direct model described in subsection \ref{subsec:Direct-model}.
Among the initial inputs of this model are the optical indexes of
the species constitution the surface. Optical indexes for ices are
documented in the literature \citep{Schmitt1998}, but are unknown
for Martian dust. Therefore, we estimated them from spectral measurement,
as described in subsection \ref{subsec:Optical-indexes-of-dust}.
Then the direct model and its inversion will be described. The final
step is the inversion method used to get information from the comparison
of our synthetic spectra with real data, detailed in the last section
\ref{subsec:Inversion}.

\subsection{Direct model\label{subsec:Direct-model}}

\paragraph*{Granular ice}

The radiative transfer model used to simulate the spectra in the case
of a translucent ice is semi analytic \citep{Doute1998}, as used
to determine seasonal cap properties \citep{Appere2011}. The surface
is supposed to be constituted of a semi infinite layer (i.e. optically
thick) that represents the CO$_{2}$ deposit, contaminated with water
ice and dust. The model uses as initial input: (i) the optical constants
used for the CO$_{2}$ and water ice are from \citep{Schmitt1998},
and the optical constants of the regolith were estimated from summer
measurements (see previous section) (ii) the grain-sizes (iii) the
proportions of each type of impurity present in the ice layer.

\paragraph*{Translucent ice}

The radiative transfer model used to simulate the spectra in the case
of a translucent ice is semi analytic \citep{Andrieu2015}. The model
uses as initial input: (i) the optical constants of the materials
constituting the surface: CO$_{2}$ and water ice from laboratory
measurements \citep{Schmitt1998}, and dust from previous consideration,
(ii) the thickness of the upper ice layer, (iii) the roughness of
the upper surface, (iv) the grain-sizes and (v) the proportions of
each type of impurity present in the ice layer and (vi) the single
scattering albedo of the underlying regolith, that is supposed to
be a semi-infinite and lambertian. The number of impurity types that
can be introduce in the model is not limited in theory, but practically,
it is not possible to introduce more than two different types due
to computation limitations.

\paragraph*{Linear mixture of defrosted terrain}

\begin{figure}
\begin{centering}
\subfloat[]{\centering{}\includegraphics[width=0.5\textwidth]{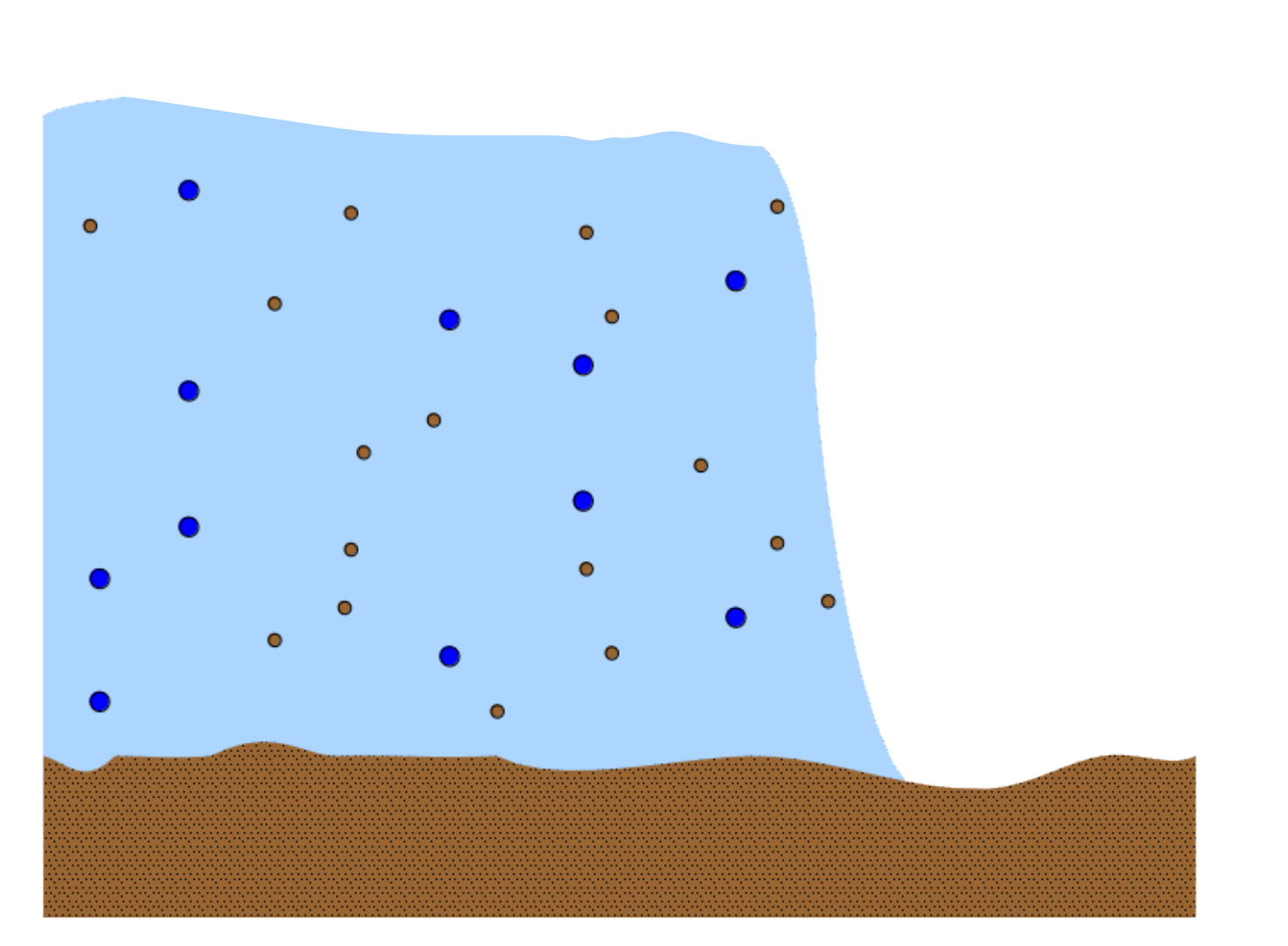}}\subfloat[]{\centering{}\includegraphics[width=0.5\textwidth]{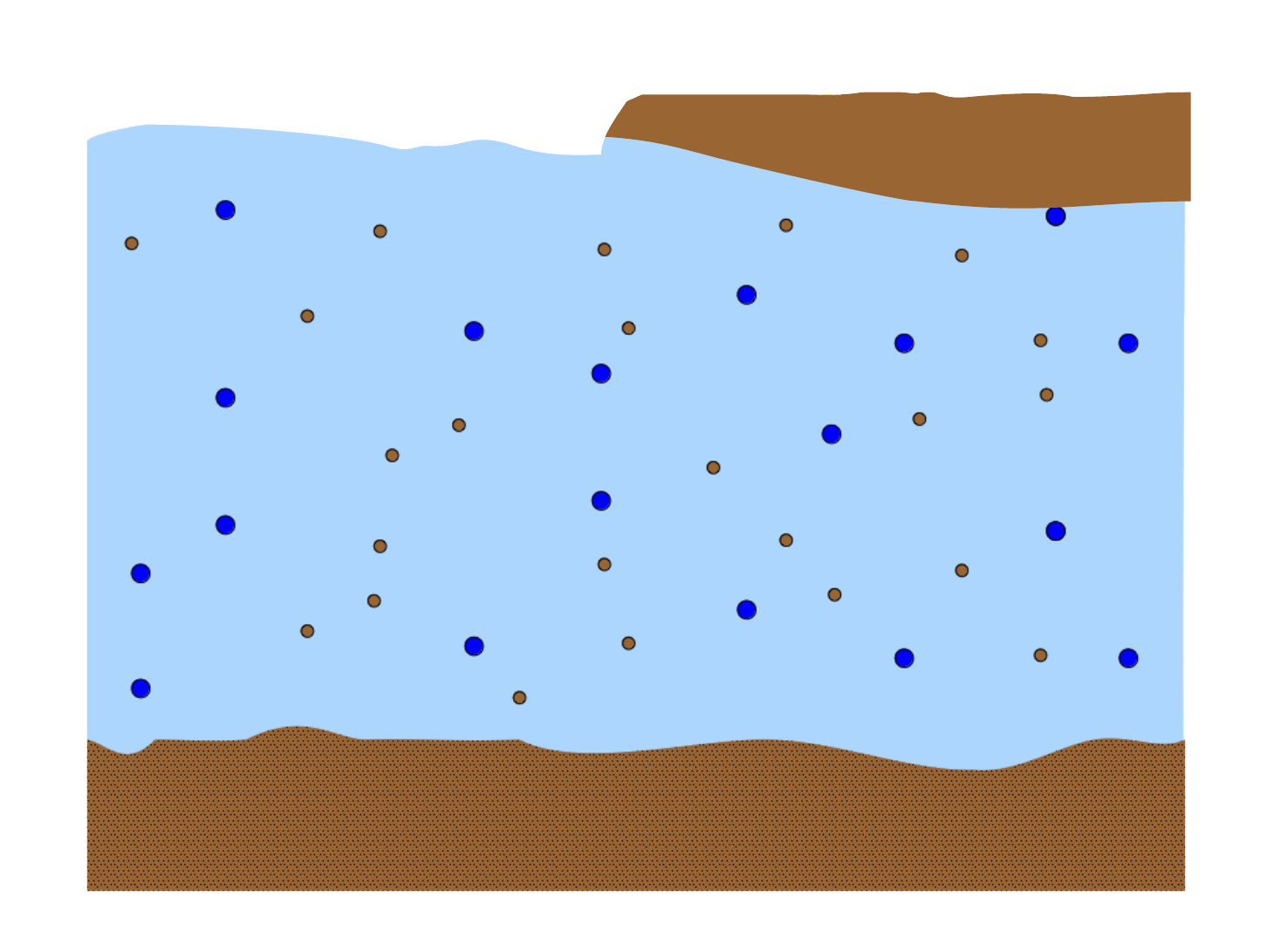}}
\par\end{centering}
\centering{}\caption{\foreignlanguage{french}{\label{fig:Surface_Model-1}\foreignlanguage{english}{Model of the
surface: a slab of CO$_{2}$ ice, possibly contaminated with water
ice and regolith dust lays on the top of the lambertian regolith.
A sub-pixel linear mixing of two different surface is possible to
model (a) the possibility of completely defrosted patches within a
pixel or (b) an optically thick layer of dust covering the ice. Both
situations result in the same modeled spectrum. }}}
\end{figure}

In order to take into account the presence of defrosted area within
one CRISM pixel, we allow a linear mixture of regolith with the previous
ice model (see fig. \ref{fig:Surface_Model-1}). Thus, one extra parameter
will be estimated: the surface proportion of naked terrain. This possibility
of sub-pixel geographical mixing appears even at the spatial resolution
of the CRISM instrument, of about 20 m per pixel, if a mixture occurs
within the same pixel. It is also possible that, following ejections,
an optically thick layer of dust covers the layer of ice locally (see
Figure \ref{fig:Surface_Model-1}). These two possibilities are described
in exactly the same way, considering a linear mixture between a surface
described by the radiative transfer model \citep{Andrieu2015}, and
a surface consisting only of regolith.

\subsection{Optical indexes of the Martian dust\label{subsec:Optical-indexes-of-dust}}

The martian southern dust differ from the northern one and from regular
palagonite \citep{Doute2007}. We may estimate the optical indexes
analyzing a defrosted area, with some assumptions. Contrary to this
previous work, we did not assume a lambertian surface for the dust
and correct for the photometric effect in order to estimate the reflectance
in a normal geometry.

\subsubsection{Single scattering albedo of the regolith}

Using the Hapke reflectance model \citep{Hapke1984}, we inverted
reflectance spectra of the ice free surface measured by the CRISM
instrument to estimate the single scattering albedo $\omega(\lambda)$,
that can be easily converted into normal reflectance. The goal is
to find the parameter $\omega(\lambda)$, assuming the other fixed
parameters, taking the average martian photometric parameters \LyXZeroWidthSpace \LyXZeroWidthSpace given
by \citealp{Vincendon2013}, \emph{i.e.} $\theta$ = 17, c = 0.6,
b = 0.12, $B_{0}$ = 1 and h = 0.05. The estimation is performed by
a bayesian inversion, for which we consider each wavelength independently
\citep{Fernando2013,Fernando2015,Schmidt2015}. The solution is a
probability density function (PDF) of $\omega$ that can be reduced
by its mean $\left\langle \omega\right\rangle $, dependent on the
wavelength.

The measured spectra data can be modeled with these photometric parameters
and with a relatively small uncertainties on $\omega$ (less than
0.05). Moreover, it can be noted in the figure \ref{fig:kn-map-1}
that the result is very stable as a function of the wavelength, which
is a guarantee of the quality of the inversion, because each wavelength
has been inverted independently.

\subsubsection{Optical indexes of the Martian dust}

\begin{figure}
\begin{centering}
\subfloat[]{\centering{}\includegraphics[width=0.5\textwidth]{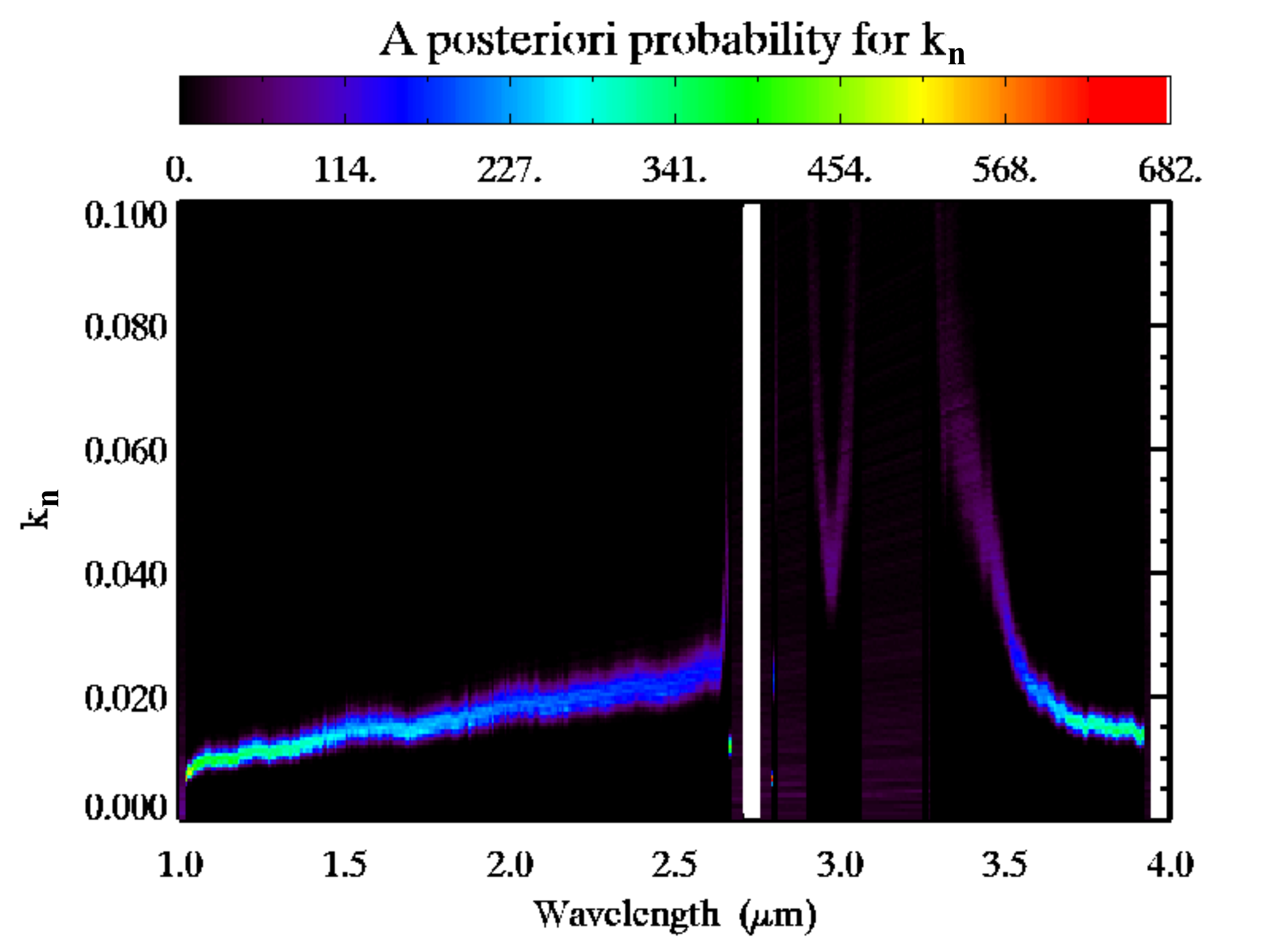}}\subfloat[]{\centering{}\includegraphics[width=0.5\textwidth]{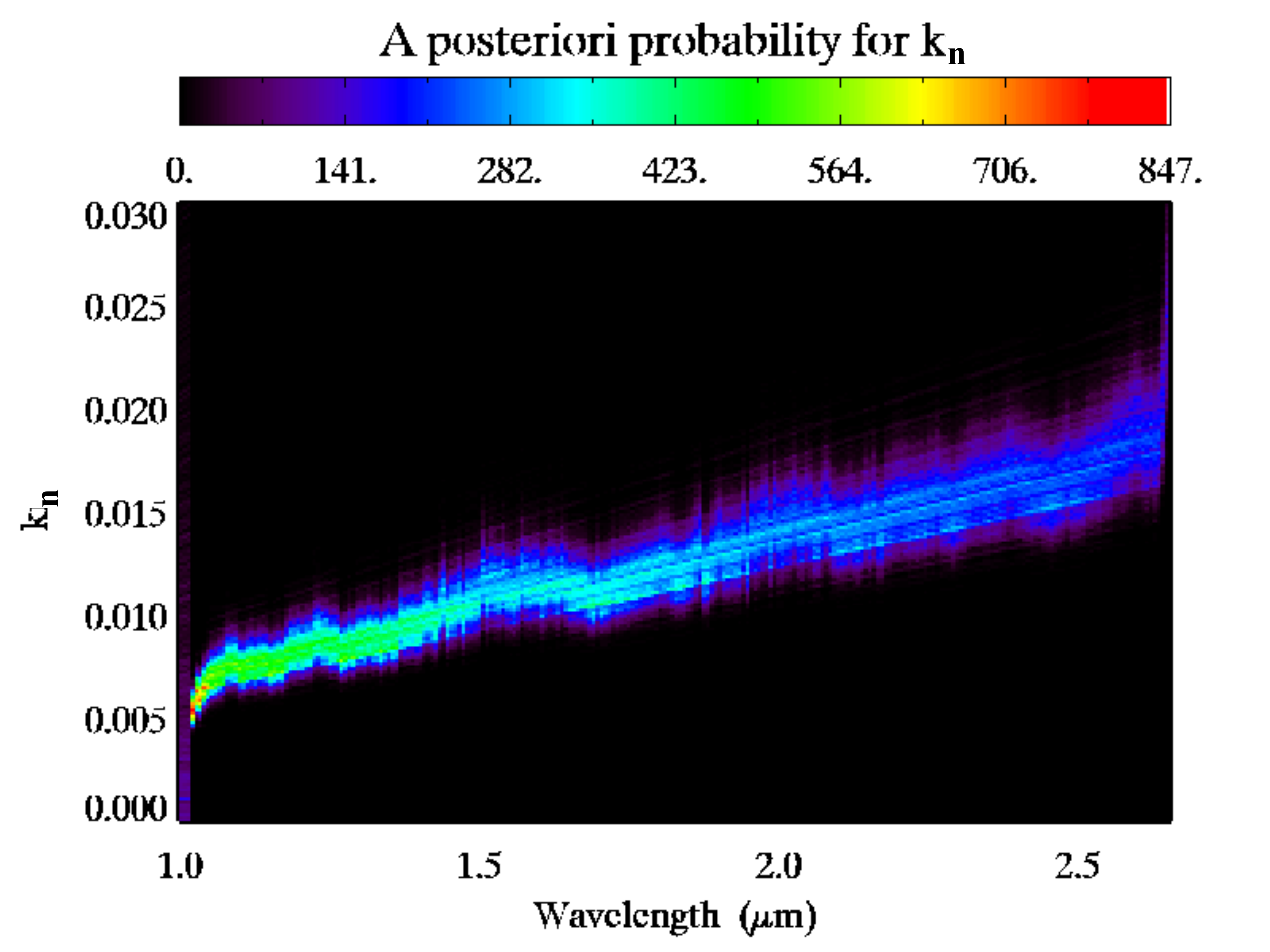}}
\par\end{centering}
\caption{\foreignlanguage{french}{\label{fig:kn-map-1}\foreignlanguage{english}{(a) Distribution of
the values \LyXZeroWidthSpace \LyXZeroWidthSpace (PDF) of the imaginary
part of the optical index $k_{n}$, calculated from the values \LyXZeroWidthSpace \LyXZeroWidthSpace of
$\omega$ of the Markov chains obtained after the Bayesian inversion,
considering a grain size of $30\,\mu$m in the Shkuratov model. We
thus propagate the distribution of the values \LyXZeroWidthSpace \LyXZeroWidthSpace of
$\omega$ on the values \LyXZeroWidthSpace \LyXZeroWidthSpace of $k_{n}$.
These values \LyXZeroWidthSpace \LyXZeroWidthSpace being determined
from the calculation of the single diffusion albedo $\omega$, the
same discontinuities are observed, and it is not possible to calculate
a value for a wavelength situated in an absorption band of the CO$_{2}$.
(b) Zoom on the part of interest located between $1\mu m$ and $2.6\mu m$
of the distribution of the values \LyXZeroWidthSpace \LyXZeroWidthSpace of
the imaginary part of the optical indices $k_{n}$, obtained from
the values \LyXZeroWidthSpace \LyXZeroWidthSpace of $\omega$ of the
Markov chains. We see that just like the distribution of $\omega$,
the distribution of the values \LyXZeroWidthSpace \LyXZeroWidthSpace of
$k_{n}$ also often admits two maxima. The value that has been retained
is the average value of this distribution.}}}
\end{figure}

The \citealp{Shkuratov1999} model is an inverse analytical model
(with an inverse analytical form) of one-dimensional geometric optics,
under normal conditions of incidence and emergence. Assuming the real
part of indexes, the size of the grains, the porosity of the surface
and knowing the value of the reflectance from the previous step, we
estimate the imaginary part $k_{n}$ of the optical index. We fix
the value for porosity of the surface at 0.5, which is the classical
value for this type of surface \citep{Hapkebook}. For the real part
of its optical index and the size of its grains, we know the order
of magnitude, but not the precise values. We have generated optical
indexes by assuming grain sizes varying from $10\,\mbox{\ensuremath{\mu}m}$
to $30\mbox{\ensuremath{\mu}m}$ and real parts of the optical indexes
ranging from 1.3 to 1.9. Using Shkuratov's inverse analytic model,
we can generate the PDF of $k_{n}$ from the PDF of reflectance (see
figure \ref{fig:kn-map-1}). There is an increase of $k_{n}$ with
the wavelength and the uncertainties are increasing, but remain relatively
low, around 0.001. The grain-size has a much bigger impact on the
form of the spectrum of $k_{n}$ than the real part of the optical
index. If we compare the values \LyXZeroWidthSpace \LyXZeroWidthSpace of
the $k_{n}$ calculated from the OMEGA \citep{Doute2007} or CRISM
data, using the same values \LyXZeroWidthSpace \LyXZeroWidthSpace for
the other parameters (real part of the optical index, grain size,
compactness), very close results are obtained, demonstrating the robustness
of the method regarding the input data. The grain size is not constrained.
Since each grain size has coherent optical index, it is impossible
to distinguish those two parameters. We choose a grain size of $30\,\mu$m
arbitrarily. It is close to the typical martian dust \citep{kieffer2007cold,Vincendon2009,Forward2009}.
In the following the retrieved dust content will be subject to this
indetermination.

\subsection{Inversion\label{subsec:Inversion}}

The inversion methods has been validated in a previous study \citep{Andrieu2016}.
The principle is to build a spectral lookup table using the radiative
transfer model, and to find a solution to the inverse problem using
bayesian statistics. The solution is a probability density function
(PDF) that is defined in the parameters space. The likelihood $L$
is estimated for each spectrum in the look up table (LUT), with:

\begin{equation}
L=\exp\left(-\frac{1}{2}\times{}^{t}\left(d_{sim}-d_{mes}\right)\overline{\overline{C}}^{-1}\left(d_{sim}-d_{mes}\right)\right)\label{eq:L-1-1-1-1}
\end{equation}
where $d_{sim}$ is a simulated spectrum from the LUT, $d_{mes}$
is the data, $^{t}$ is the transpose operator that applies to $\left(d_{sim}-d_{mes}\right)$,
and $\overline{\overline{C}}$ is the covariance matrix that represent
the uncertainties on $d_{mes}$. $d_{sim}$ and $d_{mes}$ are $n_{b}$
elements vectors corresponding to respectively simulation data and
measurements data, with $n_{b}$ the number of spectral bands used
($n_{b}=247$). 

Therefore, $\overline{\overline{C}}$ is a $n_{b}\times n_{b}$ matrix,
where each eigenvector represents a source of uncertainty, and the
associated eigenvalue represents the level of uncertainty. Usually
this matrix is assumed to be diagonal with the noise level variance
on the diagonal element. In our case, we introduced the uncertainties
due to the atmospheric correction: the shift and general slope that
may be introduced by variations of geometry at the surface, accentuated
by the atmospherical corrections (see Fig. \ref{fig:Correction-atmo}).
We represent it by two eigenvectors $\vec{t}=\frac{1}{n_{b}}\left(1,...,1\right)$
and $\vec{s}=\frac{\vec{S}}{\left\Vert \vec{S}\right\Vert }$, with
$\vec{S}=\left(-\frac{n_{b}-1}{2},\ldots,-2,-1,0,1,2,\ldots,\frac{n_{b}-1}{2}\right)$.
$\vec{s}$ characterizes the spectral slope value, and $\vec{t}$
characterizes the ordinate at the spectral shift. As $\overline{\overline{C}}$
must be a positive definite matrix, $\vec{s}$ and $\vec{t}$ must
be orthogonal, so the formulation of $\vec{s}$ is imposed by $\vec{t}$.
The 245 other eigenvectors represent the noise, and are build using
the Gram-Schmidt process. In the basis constituted of the eigenvectors,
$\overline{\overline{C}}$ is diagonal and its diagonal elements are
the eigenvalues $\sigma_{1}^{2}$, $\sigma_{2}^{2}$ and 245 times
$\sigma_{3}^{2}$ , where $\sigma_{1}$ and $\sigma_{2}$ respectively
represent a standard deviation associated to $\vec{t}$, $\vec{s}$,
and $\sigma_{3}$ represent the standard deviation associated to the
noise. We chose $\sigma_{1}=\sigma_{2}=0.02$ and $\sigma_{3}=0.03$,
in agreement with our analysis.

The likelihood computed are then converted to \emph{a} \emph{posteriori
}probability density functions (PDF). For the parameter $m_{i}$ of
the radiative transfer model, the a posteriori PDF is:

\begin{equation}
\mathcal{P}\left\{ m_{i}\right\} =\frac{\mbox{L}\left(m_{i}\right)\mbox{d}p\left(m_{i}\right)}{\sum_{i}\mbox{L}\left(m_{i}\right)\mbox{d}p\left(m_{i}\right)}
\end{equation}
where $\mbox{d}p\left(m_{i}\right)$ is the differential element at
the sampling point $m_{i}$ of the parameters space $M$: $\mbox{d}p\left(m_{i}\right)=\prod_{j}\partial p_{j}\left(m_{i}\right)$.
To make the solution easier to read, marginal a posteriori PDF are
built: the PDF is projected on one parameter's axis. This allow the
visualization of the solution for a given parameter $p_{j}$ in a
graph. For the parameter $p_{j}$, there are $n_{j}$ sampling steps
that are represented by the letter $k$ ($k$ in Eq. \ref{eq:Marginal_posterior_PDF-1}
ranges from $0$ to $n_{j}$). The marginal a posteriori PDF for $p_{j}$
at the sampling step $k$ is: 
\begin{equation}
\mathcal{P}\left\{ p_{j}(k)\right\} =\frac{\mbox{L}'\left(k\right)}{\sum_{i}\mbox{L}^{\prime}\left(k\right)\partial p_{j}\left(k\right)}\label{eq:Marginal_posterior_PDF-1}
\end{equation}
with $\partial p_{j}\left(k\right)$ the differential element for
$p_{j}$ at $k$, and: 
\begin{equation}
\mbox{L}'\left(k\right)=\sum_{i}\mbox{L}(m_{i}\mid p_{j}\left(k\right))\prod_{l\neq j}\partial p_{l}(m_{i}\mid p_{j}\left(k\right))
\end{equation}

\subsection{Look Up Tables (LUT)}

The LUT must make it possible to sample the space of the parameters
in an optimal way: representative way not to miss any information,
but not too voluminous to avoid making the calculations more complicated.
A series of tests was therefore carried out, traversing the space
coarsely following a logarithmic sampling to locate where in the parameters
space there is most variability. As we do not know precisely the optical
constants of the Martian regolith, it is useless to try to find the
size of the grains. To maintain maximum consistency, this parameter
was not sampled. The set of values sampled in the synthetic LUT is
summarized in the table \ref{tab:Ensemble-des-valeurs-2} for the
slab ice model, and in the table \ref{tab:Ensemble-des-valeurs-1-1}
for the granular ice model. 
\begin{table}
\begin{centering}
\begin{tabular*}{1\textwidth}{@{\extracolsep{\fill}}|>{\centering}m{0.35\textwidth}|>{\centering}m{0.4\textwidth}|>{\centering}m{0.15\textwidth}|}
\hline 
Parameter & Values sampled & Number of sampled values\tabularnewline
\hline 
\hline 
Thickness of the CO$_{2}$ slab$\left(\mbox{mm}\right)$  & $0$ ; $0.01$ ; $0.05-1$ in steps of $0.05$ ; $1-200$ in steps
of $1$ ; $200-300$ in steps of $5$ ; $300-1000$ in steps of $50$  & $250$\tabularnewline
\hline 
Total amount of impurities $\left(\mbox{\%}\right)$ & $0$ ; $0.005$ ; $0.01$ ; $0.05-1$ in steps of $0.05$ ; $10-100$
in steps of $10$ ; $10-100$ in steps of $10$ & $32$\tabularnewline
\hline 
Dust proportions among impurities $\left(\mbox{\%}\right)$ & $0$ ; $1$ ; $10$ ; $50$ ; $90$ ; $99$ ; $100$  & $7$\tabularnewline
\hline 
Water ice proportions among impurities $\left(\mbox{\%}\right)$ & $0$ ; $1$ ; $10$ ; $50$ ; $90$ ; $99$ ; $100$  & $7$\tabularnewline
\hline 
Water ice grain diameter $\left(\mbox{\ensuremath{\mu}m}\right)$ & 1 ; $3$ ; $10$ ; $30$ ; $100$; $300$ ; $1000$; $3000$ ; $10000$
; $30000$ & $10$\tabularnewline
\hline 
Surface proportion of dust in linear mixing $\left(\mbox{\%}\right)$  & $0-100$ in steps of $10$ & $11$\tabularnewline
\hline 
Incidence & $55\text{\textdegree}-80\text{\textdegree}$ in steps of $5\text{\textdegree}$ & $6$\tabularnewline
\hline 
Emergence & $0\text{\textdegree}-30\text{\textdegree}$ in steps of $10\text{\textdegree}$ & $4$\tabularnewline
\hline 
Azimuth & $60\text{\textdegree},150\text{\textdegree}$ & $2$\tabularnewline
\hline 
\end{tabular*}
\par\end{centering}
\caption{\label{tab:Ensemble-des-valeurs-2}Values sampled in the LUT for the
translucent slab model.}
\end{table}
 
\begin{table}
\begin{centering}
\begin{tabular*}{1\textwidth}{@{\extracolsep{\fill}}|>{\centering}m{0.35\textwidth}|>{\centering}m{0.4\textwidth}|>{\centering}m{0.15\textwidth}|}
\hline 
Parameter & Values sampled & Number of sampled values\tabularnewline
\hline 
\hline 
Grain diameter of CO$_{2}$ ice $\left(\mbox{\ensuremath{\mu}m}\right)$  & $1$ ; $3$ ; $10$ ; $30$ ; $100$ ; $300$ ; $1000$ ; $3000$
; $10000$ ; $30000$ ; $100000$ ; $300000$ & $12$\tabularnewline
\hline 
Total amount of impurities $\left(\mbox{\%}\right)$ & $0$ ; $0.005$ ; $0.01$ ; $0.05-1$ par pas de $0.05$ ; $10-100$
in steps of $10$ ; $10-100$ in steps of $10$ & $32$\tabularnewline
\hline 
Dust proportions among impurities $\left(\mbox{\%}\right)$ & $0$ ; $1$ ; $10$ ; $50$ ; $90$ ; $99$ ; $100$  & $7$\tabularnewline
\hline 
Water ice proportions among impurities $\left(\mbox{\%}\right)$ & $0$ ; $1$ ; $10$ ; $50$ ; $90$ ; $99$ ; $100$  & $7$\tabularnewline
\hline 
Water ice grain diameter $\left(\mbox{\ensuremath{\mu}m}\right)$ & 1 ; $3$ ; $10$ ; $30$ ; $100$; $300$ ; $1000$; $3000$ ; $10000$
; $30000$ & $10$\tabularnewline
\hline 
Surface proportion of dust in linear mixing $\left(\mbox{\%}\right)$  & $0-100$ in steps of $10$ & $11$\tabularnewline
\hline 
Incidence & $55\text{\textdegree}-80\text{\textdegree}$ in steps of $5\text{\textdegree}$ & $6$\tabularnewline
\hline 
Emergence & $0\text{\textdegree}-30\text{\textdegree}$ in steps of $10\text{\textdegree}$ & $4$\tabularnewline
\hline 
Azimuth & $60\text{\textdegree},150\text{\textdegree}$ & $2$\tabularnewline
\hline 
\end{tabular*}
\par\end{centering}
\caption{\label{tab:Ensemble-des-valeurs-1-1}Values sampled in the LUT for
the compact granular layer model \citeauthor{Doute1998}.}
\end{table}

This synthetic LUT for the slab model contains $9939264$ spectra,
corresponding to $207068$ configurations of different surfaces for
$48$ geometries of measurements, representing the variability of
the direct model. Each spectrum was calculated at high spectral resolution
($29037$ wavelengths at a spectral resolution of about $0.5\text{\,cm}^{-1}$),
and then resampled at the resolution of the CRISM instrument ($247$
wavelengths used here, at a resolution of $6.5\,\text{nm}$), using
the transfer function of the sensor. It was computed in $60\,\text{h}$,
in parallel on four Intel Xeon CPU cores Xeon E5640, $2.66\,\text{GHz}$.

To take the sub-pixel mixture into account during the inversion, the
LUT is extended by calculating the mixture with a martian dust spectrum
at the corresponding geometry and for the different desired surface
ratios. In this study, we sampled $11$ values of surface proportions
of linear mixing with dust, ranging from $0
$ to $100
$ in steps of $10
$. The LUT compared to the data during the inversion is thus $11$
times more voluminous than that described previously, and therefore
contains a total of $109331904$ spectra, corresponding to $2277748$
different surface configurations.

\section{Data}

\subsection{CRISM data}

The CRISM instrument \citep{Murchie2007} has well covered the Richardson
Crater area, and provide 12 cubes of the same area for the spring
martian year 28. This time coverage is enough to follow the general
time evolution of seasonal deposits through spring: the mean time
span between two acquisitions is 9\textdegree{} of solar longitude,
or 16.7 \emph{sols}. Another advantage of the CRISM data is that most
of the observations are coupled with HiRISE images \citep{McEwen2007}
of the same area, taken at the same moment, at a resolution of 25cm/px.
This allows to investigate closely the relations between spectroscopic
and spatial informations. In particular, the dark dune spots, widely
active in the area \citep{Kereszturi2011} have a spatial extension
of a few CRISM pixels, and thus their evolution can be monitored both
spectroscopically using CRISM and geomorphologically using HiRISE
data. 

The purpose of this work was to investigate precisely the interactions
between surface and atmosphere during the spring defrosting in areas
that show defrosting spring activity. These spring defrosting manifestations
have a typical scale of a few hundreds meters to a few kilometers.
The resolution of the instrument used to study them shall then be
lower than a hundred meters. We thus chose the instrument CRISM that
provide the best possible spatial resolution (approx. 20m per pixel)
compared to others imaging spectrometers. Moreover, CRISM acquires
data always at the same local solar time, so diurnal variations are
not to be taken into account.

\begin{figure}
\begin{centering}
\subfloat[]{\includegraphics[height=7cm]{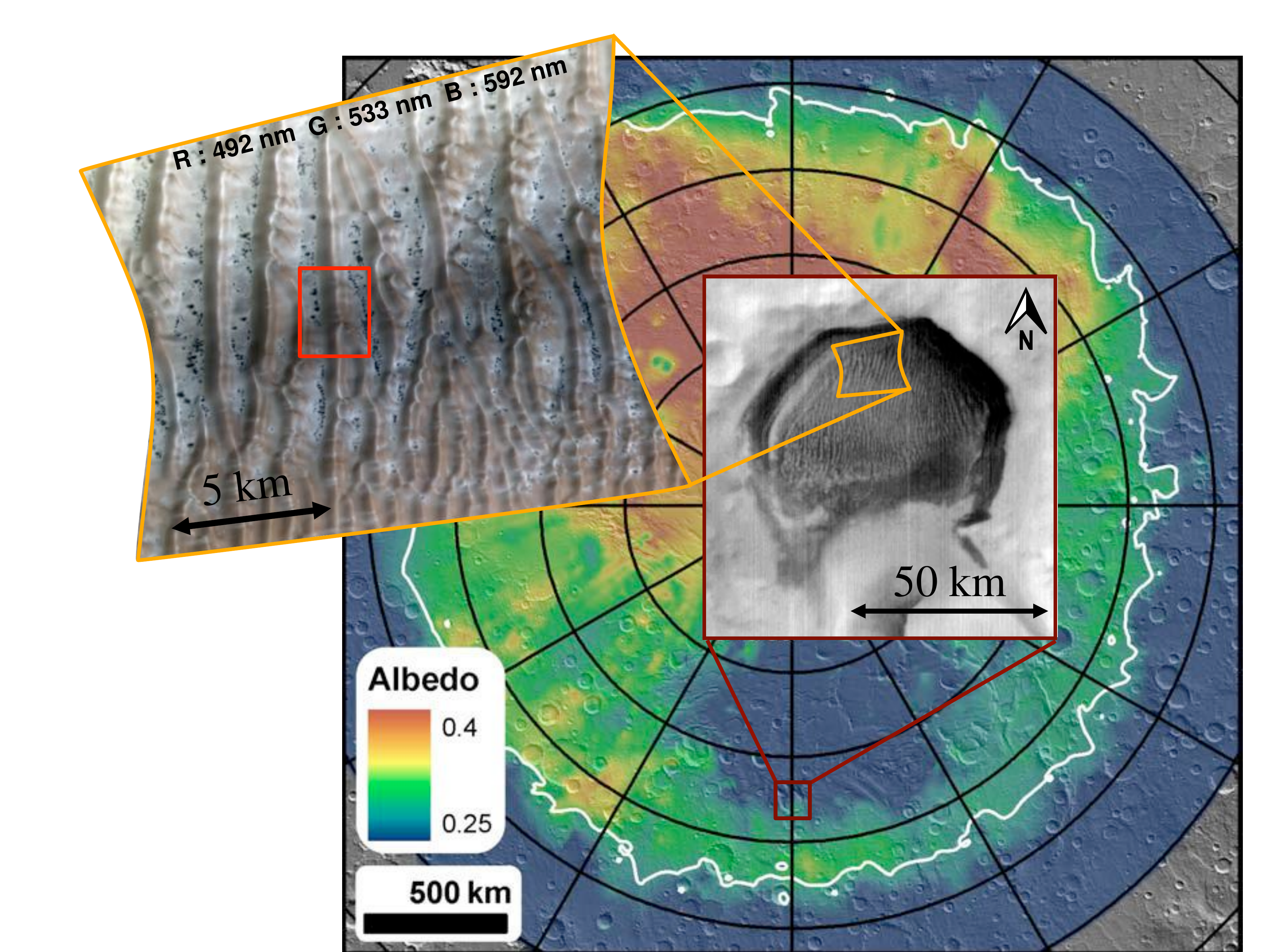}}\,\subfloat[]{\includegraphics[height=6.5cm]{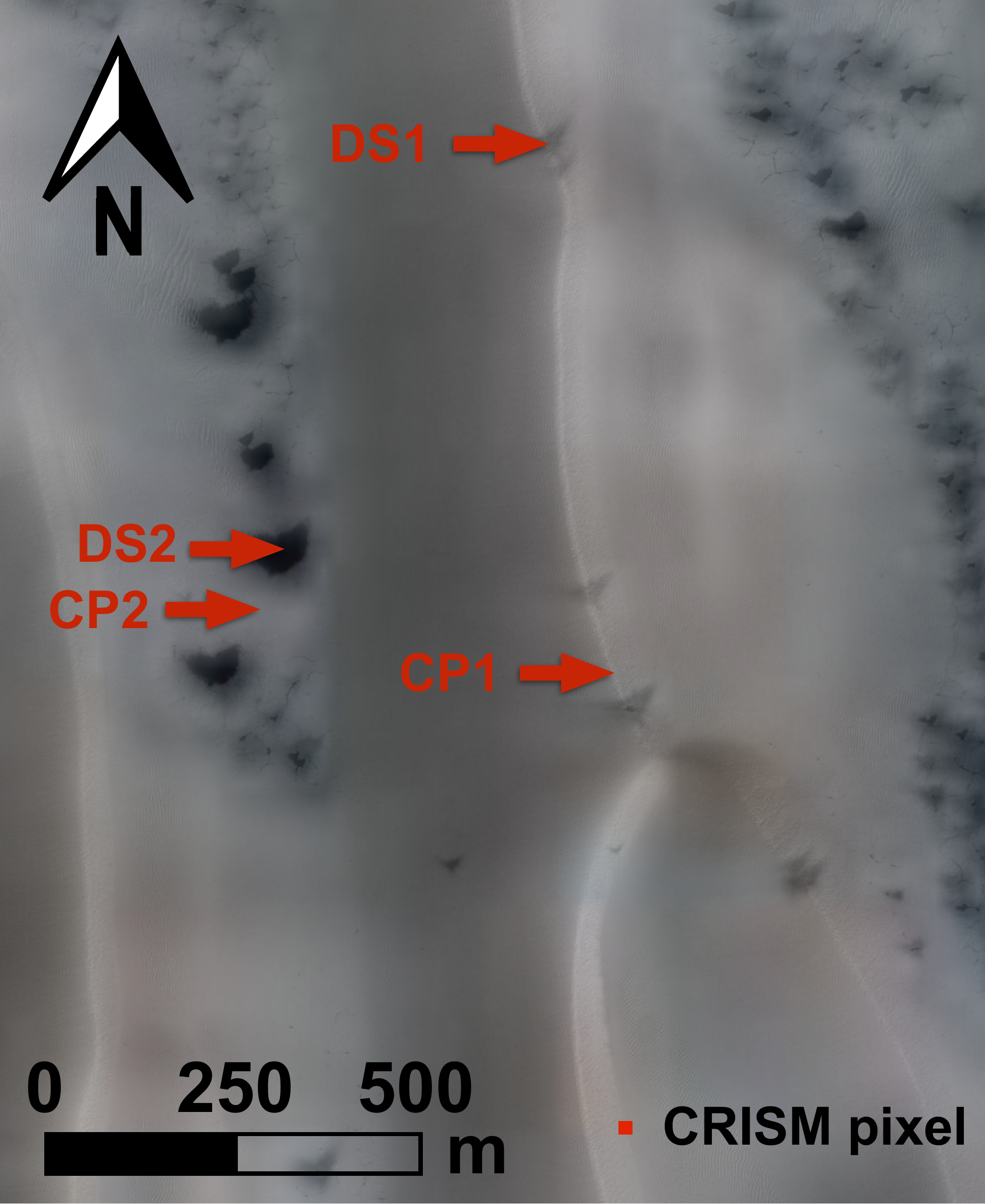}

}
\par\end{centering}
\caption{\label{fig:Context_map}(a) localization of Richardson crater, Mars.
The back is a polar projection map centered on the south pole of Mars,
from \citep{Hansen2010}, where the limit of the seasonal CO$_{2}$
ice deposits in the middle of the spring (220\textdegree < Ls<225\textdegree )
is represented in white. A MOC image of Richardson crater, located
at 72.6\textdegree S, 180.4\textdegree W, is superimposed, and then
a map projected color composition of a CRISM cube FRT000052BC (LS=213\textdegree ).
(b) Detail of areas of particular interest: extract from the CRISM
color composition in (a) superimposed on an extract from HiRISE image
PSP\_002542\_1080. Two control points (CP1 and CP2) and two points
of particular interest that show defrosting activity (DS1 and DS2)
were defined.}
\end{figure}

We searched for areas of particular interest, that presented a time
coverage by CRISM that was sufficient to follow the evolution of the
seasonal deposits. We chose the Richardson crater dune field (See
Figure \ref{fig:Context_map}) for the following reasons: 
\begin{enumerate}
\item The time sampling by CRISM was sufficient for the spring of martian
year 28 (12 data cubes available).
\item The area is fully covered by CO$_{2}$ ice at the beginning of the
time series and fully defrosted at the end.
\item This area showed an intense surface process activity.
\item Several previous studies demonstrated the abundant presence of water
ice in this area, suggesting possible interactions between water and
CO$_{2}$ cycles, and justifying the use of complex radiative transfer
simulations to model the radiation in the surface. 
\end{enumerate}
\begin{table}
\centering{}%
\begin{tabular}{>{\centering}p{2.5cm}>{\centering}p{2cm}>{\centering}p{1.5cm}>{\centering}p{1.5cm}>{\centering}p{2.5cm}>{\centering}m{3cm}}
\hline 
Cube ID & Acquisition date & $L_{S}\,\left(\text{\textdegree}\right)$ & AOT & Mean solar incidence $\left(\text{\textdegree}\right)$ & Comments\tabularnewline
\hline 
\noalign{\vskip\doublerulesep}
\hline 
HRS0000411F & 30/01/2007 & $175.213$ & n. c. & $83.9$ & Atmosphere saturated\tabularnewline
\noalign{\vskip\doublerulesep}
\hline 
HRL000043A2 & 10/02/2007 & $181.550$ & $0.59$ & $80.9$ & CO$_{2}$ and water ice at the surface\tabularnewline
\noalign{\vskip\doublerulesep}
\hline 
HRS000049E5 & 09/03/2007 & $197.016$ & $1.11$ & $75.9$ & CO$_{2}$ and water ice at the surface\tabularnewline
\noalign{\vskip\doublerulesep}
\hline 
FRT000052BC & 05/04/2007 & $213.719$ & $0.92$ & $69.1$ & CO$_{2}$ and water ice at the surface\tabularnewline
\noalign{\vskip\doublerulesep}
\hline 
FRT000054E5 & 12/04/2007 & $217.503$ & $0.72$ & $66.7$ & CO$_{2}$ and water ice at the surface\tabularnewline
\noalign{\vskip\doublerulesep}
\hline 
HRS000056C0 & 17/04/2007 & $220.683$ & $0.43$ & $66.1$ & CO$_{2}$ and water ice at the surface\tabularnewline
\noalign{\vskip\doublerulesep}
\hline 
FRT00005AF7 & 15/05/2007 & $238.066$ & $1.32$ & $60.2$ & CO$_{2}$ and water ice at the surface\tabularnewline
\noalign{\vskip\doublerulesep}
\hline 
FRT00005C94 & 20/05/2007 & $241.319$ & n. c. & $59.8$ & CO$_{2}$ and water ice at the surface\tabularnewline
\noalign{\vskip\doublerulesep}
\hline 
FRT00005E38 & 26/05/2007 & $245.220$ & $1.17$ & $57.9$ & CO$_{2}$ and water ice at the surface\tabularnewline
\noalign{\vskip\doublerulesep}
\hline 
FRT00005FF6 & 31/05/2007 & $248.483$ & $1.32$ & $57.6$ & CO$_{2}$ and water ice at the surface\tabularnewline
\noalign{\vskip\doublerulesep}
\hline 
FRT00006102 & 04/06/2007 & $251.104$ & $1.51$ & $59.0$ & No CO$_{2}$ ice at the surface\tabularnewline
\noalign{\vskip\doublerulesep}
\hline 
FRT00006C56 & 25/07/2007 & $283.403$ & $2.1$ & $53.7$ & No CO$_{2}$ ice at the surface\tabularnewline
\hline 
\noalign{\vskip\doublerulesep}
\end{tabular}\caption{\foreignlanguage{french}{\label{tab:R=0000E9capitulatif-des-observations}\foreignlanguage{english}{Summary
of available CRISM data for Richardson crater dune field, and for
martian year 28. Dust aerosol optical thicknesses (AOT) are given
by M. J. Wolff \citep{Wolff2009}. <<n. c.>> (not calculated) means
M.J. Wolff's algorithm could not retrieve the dust AOT. Aerosols effect
were not corrected in this case. Incidences are defined with respect
to the Areoide. }}}
\end{table}
The 12 available CRISM cubes for this area are summarized in table
\ref{tab:R=0000E9capitulatif-des-observations}. The first cube of
the spring (HRS0000411F) turned out to be ineffective, because a large
number of bands were saturated in atmosphere due to the high incidence
angle and because of the generally low albedo of the surface. From
Ls=251\textdegree{} and after, no CO$_{2}$ ice can be detected at
the surface using band criterion methods \citep{Langevin2007}, meaning
the the seasonal ice deposits have completely sublimed. 

\begin{figure}
\centering{}\includegraphics[width=1\paperwidth]{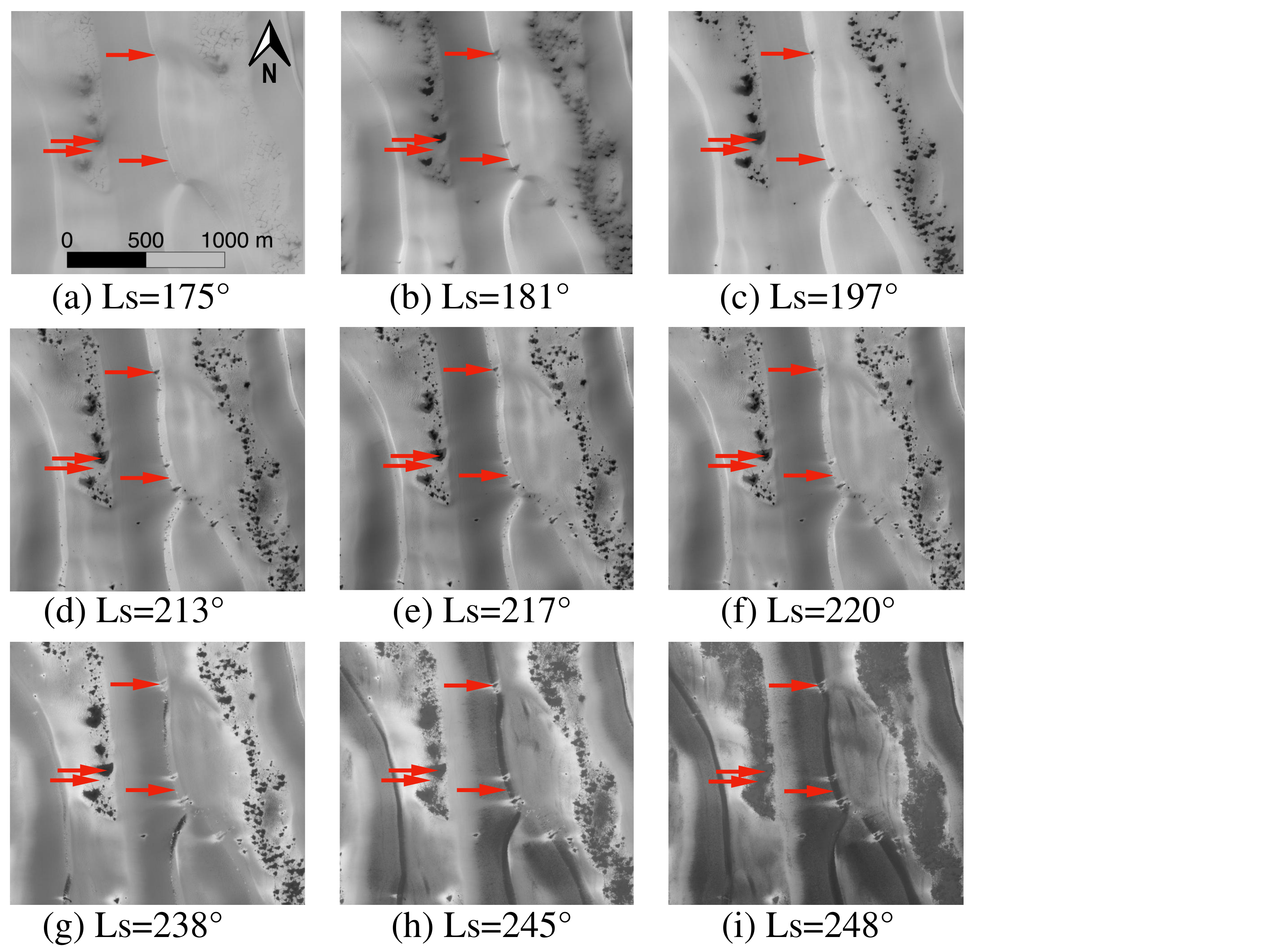}\caption{\label{fig:hirise_series}HiRISE series of the area of particular
interest defined in Figure \ref{fig:Context_map}b, during the spring
of MY28. Red arrows indicate the positions of the points CP1, CP2,
DS1 and DS2 defined in figure \ref{fig:Context_map}b. }
\end{figure}

In the area, we chose to focus on a few different spots, and to investigate
deeply their evolutions. We used the combined HiRISE and CRISM data
to select a few ``control points'', where no dark spot of flow activity
is recorded during the spring, and some point precisely where a dark
spot appears (See Figure \ref{fig:Context_map} and \ref{fig:hirise_series}
for details). The temporal series of spectra corresponding to each
study area were then extracted from the CRISM cubes. 

\subsection{Noise level}

The nominal CRISM instrumental noise is very low, and still was at
the time when the data were acquired (2007), with a signal to noise
ratio between 400 and 500 depending on the wavelength. Nevertheless,
the real data show a level far higher, most probably due to detector
electronics, atmospheric perturbations, or geometric variations within
a pixel. Computing the covariance matrix of a set of spectra that
are supposed to be close to identical allowed us to estimate that
level of noise at approximately $3\mbox{\%}$ of the signal, consistently
with previous studies \citep{Ceamanos2013,Fernando2013}.

\subsection{Atmospheric correction}

The atmospheric gas contribution was corrected using \emph{ab initio
}line by line radiative transfer calculations\citep{Doute2014}, based
on general circulation models (GCM) predictions \citep{Forget99,Forget2006}.
The mineral aerosols atmospheric contribution was corrected using
aerosols optical thickness (AOT) estimations by \citealp{Wolff2009}
and the method developed by \citealp{Doute2013}, that take into account
the radiative coupling between gas and aerosols. No correction method
has proven 100\% efficient to remove the aerosols contribution for
CO$_{2}$ ice covered areas \citep{Vincendon2007,Vincendon2008,Smith2009,Wolff2009,Ceamanos2013,Doute2013,Doute2014}.
Therefore, we expect correction errors, that have to be taken into
account in the inversions. We chose to estimate that level of error,
and to introduce it in the inversion framework as a source of uncertainty. 

\begin{figure}
\begin{centering}
\subfloat[]{\begin{centering}
\includegraphics[width=0.48\textwidth]{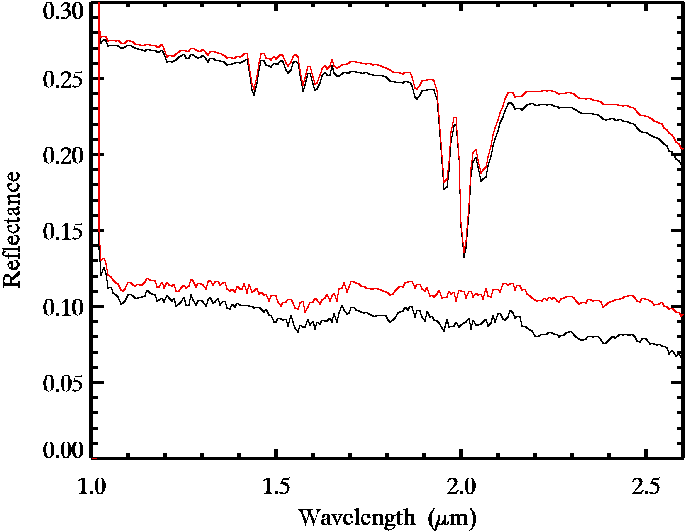}
\par\end{centering}
}\,\subfloat[]{\begin{centering}
\includegraphics[width=0.48\textwidth]{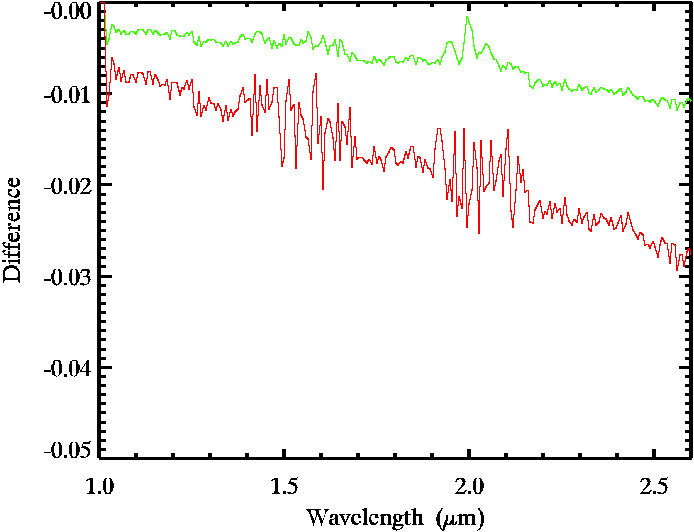}
\par\end{centering}
}
\par\end{centering}
\caption{\foreignlanguage{french}{\label{fig:Correction-atmo}\foreignlanguage{english}{Effect of local
slopes at the surface, not taken into account in the atmospheric corrections
\citep{Doute2013}. (a) Two spectra that are supposed to be identical
before (up) and after (down) atmospheric corrections. The black lines
represent the spectra before and after correction in a flat area,
and the red ones in a slope. (b) Difference between the two supposed
identical spectra (red spectra minus black spectra from fig. a), before
(green) and after (red) atmospherical correction. There is a difference
between the two supposed identical spectra, and the atmospherical
corrections seem to enhance it, by adding a general spectral slope
to the data.}}}
\end{figure}
As shown on figure \ref{fig:Correction-atmo}, the atmospheric corrections
may introduce a bias in the data in case of a sloped facet: in this
graph are represented two spectra from the same region during summer.
The whole area is homogeneously covered in dust and thus is not expected
to show any spectral variability. The variability here is correlated
with photometry, and thus attributed to a bias in atmospherical corrections.
The bias as been estimated to be a general negative spectral slope
of maximal magnitude of 0.01UR/\textmu m (UR mean units of reflectance
factor, that is dimensionless) but also a shift in the absolute level
in the order of 0.02 UR. The covariance matrix of the uncertainties
$\overline{\overline{C}}$ has to take these effects into account. 

\section{Results}

\subsection{Best fit\label{subsec:Ajustements}}

\begin{figure}
\centering{}\includegraphics[width=1\textwidth,height=10cm]{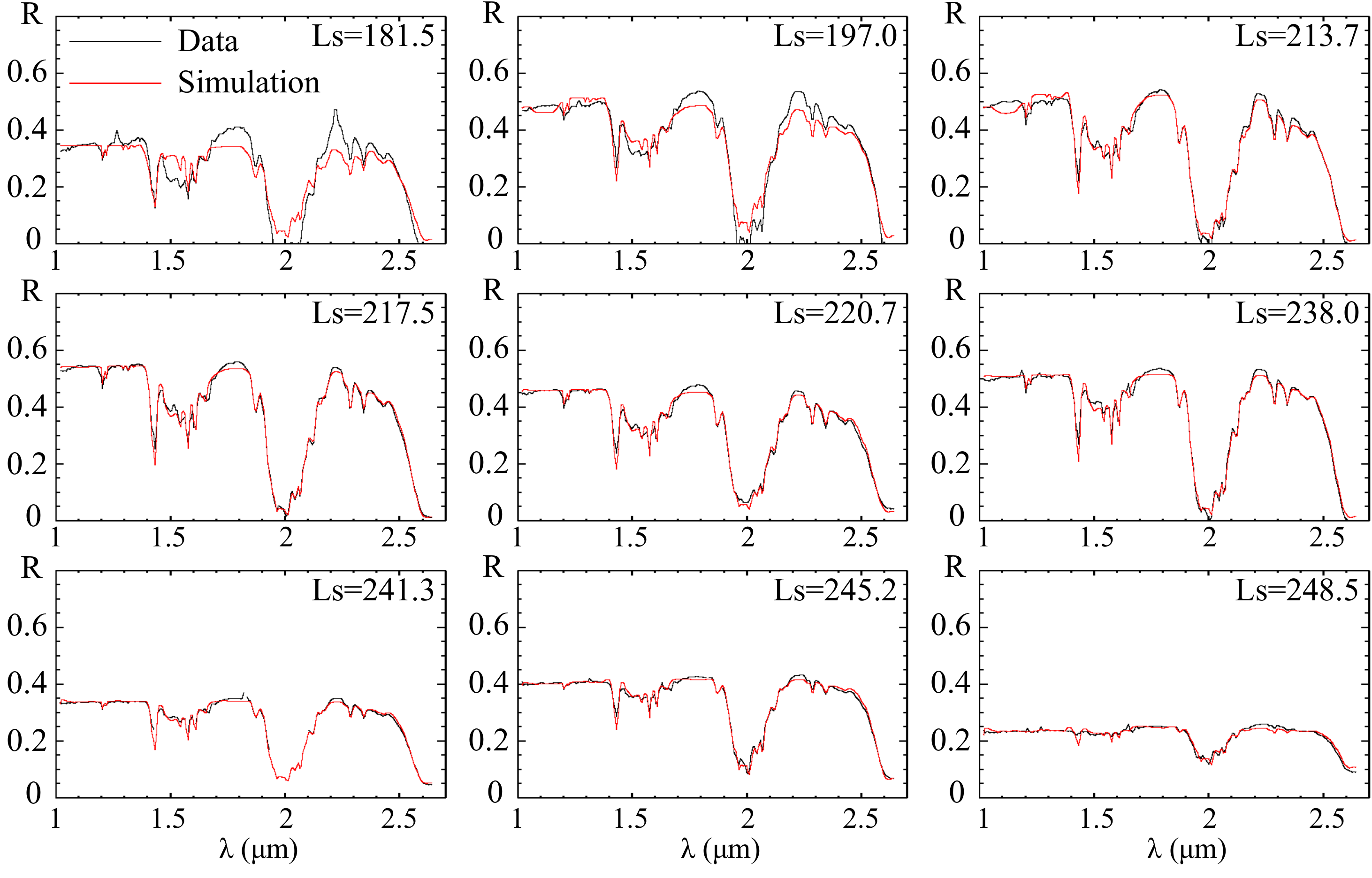}\caption{\foreignlanguage{french}{\label{fig:fits-CP1-translucent}\foreignlanguage{english}{ Best fit
of translucent ice model as a comparison with the data for temporal
series of point CP1 from $L_{S}=181.5\text{\textdegree}$ to $L_{S}=248.5\text{\textdegree}$.
The best fit spectra are re-adjusted at the level of deformation estimated
by the inversion.}}}
\end{figure}
\begin{figure}
\centering{}\includegraphics[width=1\textwidth,height=10cm]{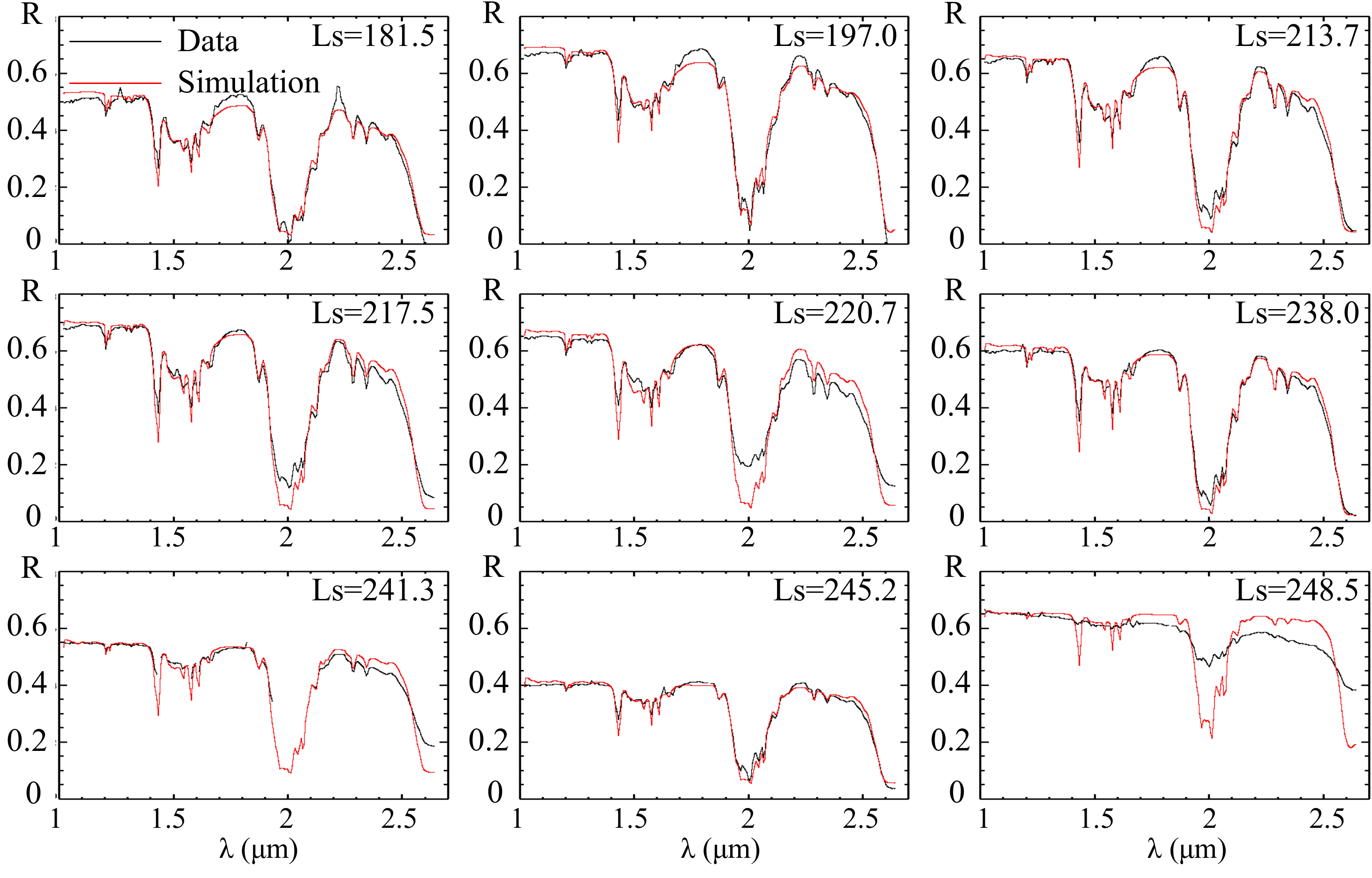}\caption{\foreignlanguage{french}{\label{fig:fits-CP1-granular}\foreignlanguage{english}{Best fit of
granular ice model as a comparison with the data for temporal series
for CP1 from $L_{S}=181.5\text{\textdegree}$ to $L_{S}=248.5\text{\textdegree}$.
The best fit spectra are re-adjusted at the level of deformation estimated
by the inversion.}}}
\end{figure}
The inversion method described above takes into account the fact that
the reflectance spectrum of the surface may be slightly deformed by
a spectral slope and level due to imperfect atmospheric correction.
It is crucial to avoid interpreting these deformations in terms of
composition and surface condition erroneously. Not only this potential
bias is ignored by the inversion algorithm thanks to the first two
eigenvectors of $\overline{\overline{C}}$ (see section~\ref{subsec:Inversion}),
but it is estimated and can therefore be corrected from the data.
The adjustments shown in the figures \ref{fig:fits-CP1-translucent}
and \ref{fig:fits-CP1-granular} were done by correcting this bias
according to the level estimated by the inversion algorithm. As demonstrated
by these figures, the best fits are usually in very good agreement
with the data. The worst is the first, due to spikes at wavelengths
around 1.7 $\mu$m and 2.2 $\mu$m in the observations that are most
probably artifacts.

\subsection{Slab vs granular ice}

\begin{figure}
\begin{centering}
\includegraphics[width=10cm]{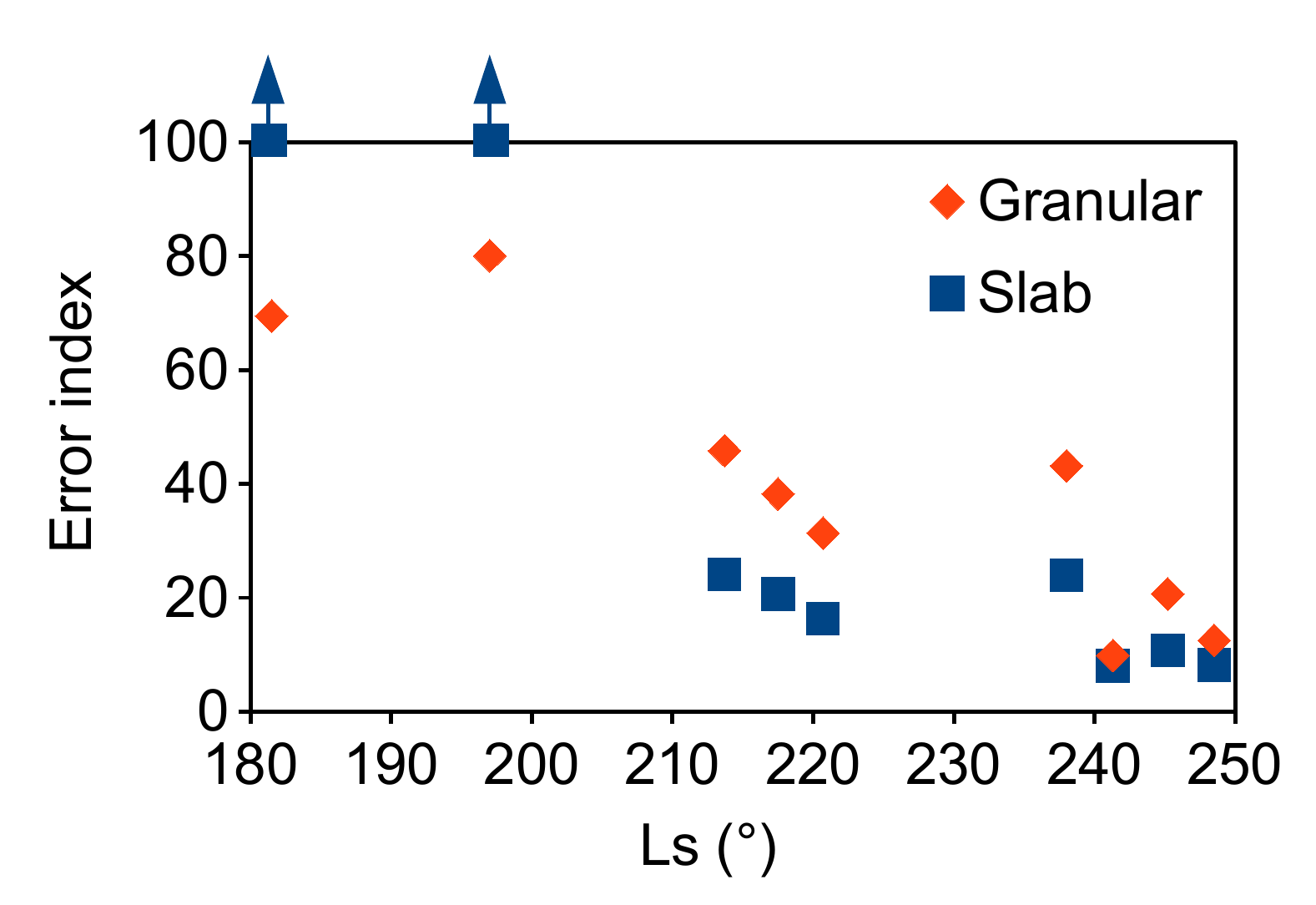}
\par\end{centering}
\caption{\label{fig:Error_index}Error indexes ($EI$) of the fits for point
CP1 for the slab and granular models}
\end{figure}

\begin{figure}
\centering{}\subfloat[Granular]{\centering{}\includegraphics[width=0.5\textwidth,height=5.5cm]{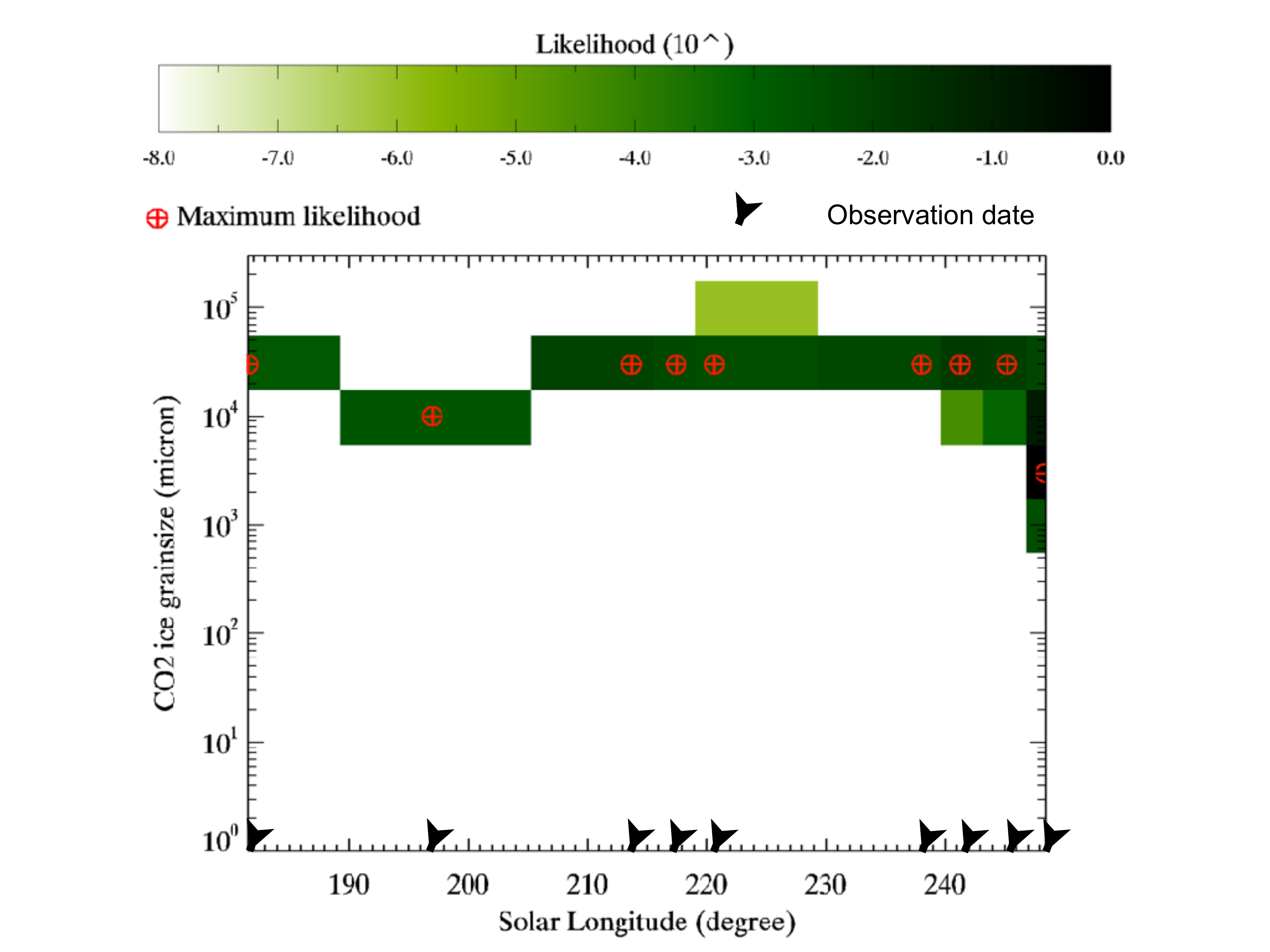}}\subfloat[Slab]{\centering{}\includegraphics[width=0.5\textwidth,height=5.5cm]{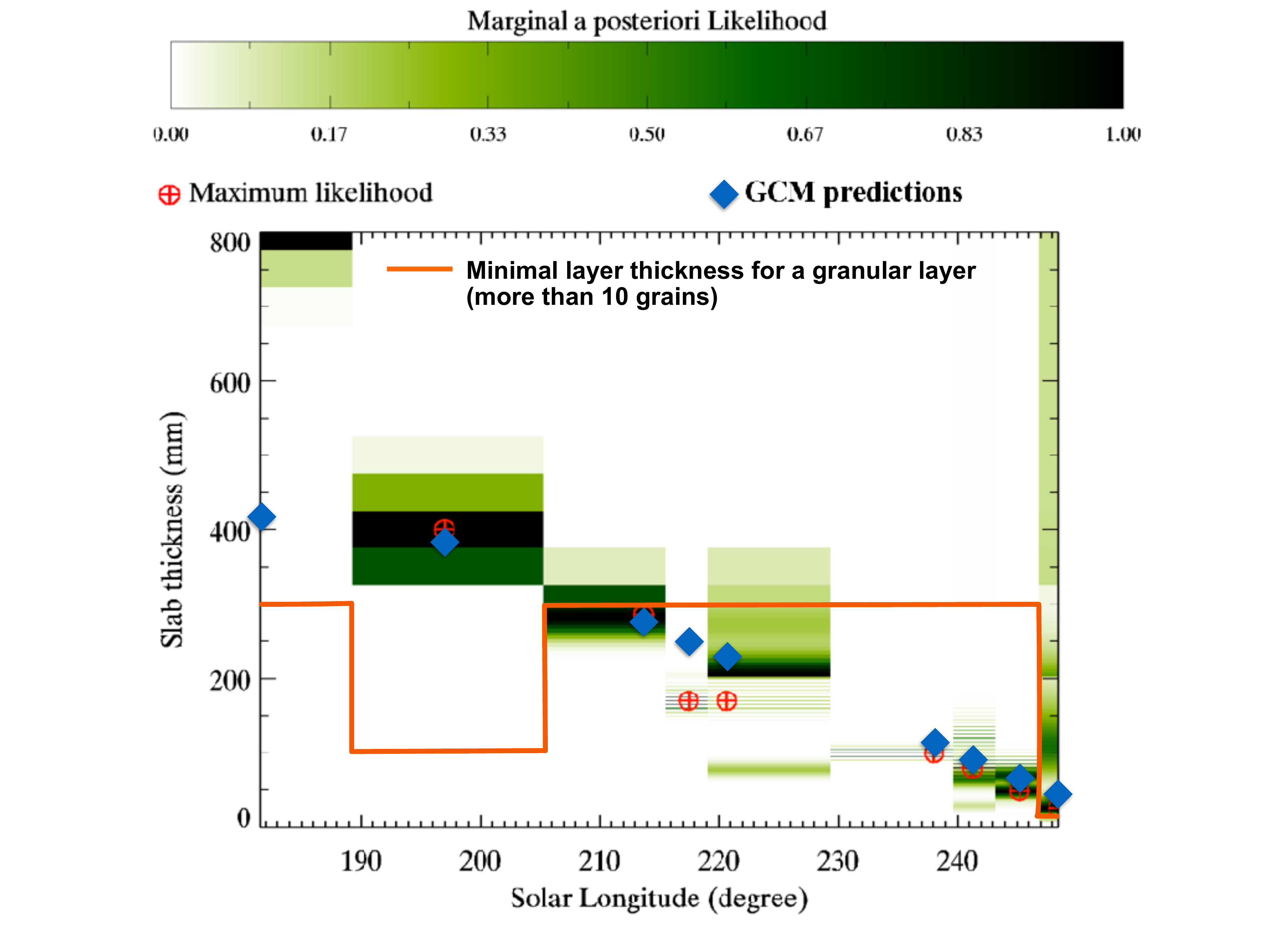}}\caption{\foreignlanguage{french}{\label{fig:granularVSslab}\foreignlanguage{english}{ (a) Evolution
of the grain size in a granular ice layer case for CP1. The shades
of green represent the values of the marginal a posteriori of the
likelihood PDF, and the red crosses point out the maximum value of
the likelihood PDF (best fit), at each observation date. (b) Evolution
of the thickness of the ice layer as a function of Ls for CP1, in
the case of a slab layer. The GCM predicted evolution (\citealp{millour2014latest}
©LMD/OU/IAA/ESA/CNES) has been added in blue diamonds, considering
a density for CO$_{2}$ ice of $\rho=1606\,\mbox{kg}\mbox{m}^{-3}$
\citep{kieffer2007cold}, and the minimal thickness possible for a
granular layer, derived from the grain-diameters given by figure a
have been added in red.}}}
\end{figure}

Both slab ice or granular ice models fit reasonably well the data.
Therefore, the quality of the fit will not be determinant to decide
which model is closer to reality. Figure \ref{fig:Error_index} shows
an error index $EI=\left|\ln L\right|$, where L is the likelihood.
Thus high value of $EI$ means bad fit, and $EI=0$ means a perfect
fit. In terms of error index, the granular model fits better (even
if badly ) for the first two measures of the season, and then the
slab model have a lower error index. Nevertheless both models show
an error index of similar magnitude. The quality of the fit then is
inappropriate to discriminate the models. 
 To test the coherence of these model, we compare them with other
independent estimates. Predictions of CO$_{2}$ ice quantity (in kg.m$^{-2}$)
at the surface by climate models are readily available via the Mars
Climate Database \footnote{http://www-mars.lmd.jussieu.fr/mcd\_python/}
\citep{Lewis99,Forget99,Forget2006,Millour2012,millour2014latest}.
The thicknesses are then deduced from the GCM prediction by considering
an homogeneous layer, and mass density of $1606$ kg.m$^{-3}$ \citep{kieffer2007cold,Thomas2011}.
The thickness data derived from the climate models are then plotted
in the graphs with ice thickness estimates in Fig. \ref{fig:granularVSslab}.
We can see in the figure \ref{fig:granularVSslab} that the direct
estimates realized with our algorithm and the predictions from the
climate models are consistent. Similar orders of magnitude, and often
very close estimates are observed, except for Ls = 181.5\textdegree ,
where we observe thicknesses clearly greater than those predicted. 

For the granular ice model, we can estimate the minimum thickness
of granular ice required for a optically thick layer. Since the ice
is poorly absorbent, a layer of 10 grains is required to fulfill this
property (see fig. \ref{fig:granularVSslab}). We can thus compute
the minimum thickness in the case of a granular ice that is inconsistent
with the MCD, even with the expected behavior of sublimating. The
grain diameter in the case of a granular model shows no clear evolution
during spring (see figure \ref{fig:granularVSslab}a), and figure
\ref{fig:granularVSslab}b shows that the minimum thickness required
to fulfill the granular model assumptions is highly inconsistent with
the climate models predictions. On the contrary, the thickness estimates
from the slab ice models show a decrease during spring in agreement
with the values predicted by the Mars Climate Database. From this
result, the granular ice can be ruled out and be ignored for the rest
of this article.

\subsection{Thickness\label{subsec:=0000C9paisseur}}

\begin{figure}
\centering{}\subfloat[\foreignlanguage{french}{Point CP1}]{\centering{}\includegraphics[width=0.5\textwidth,height=5.5cm]{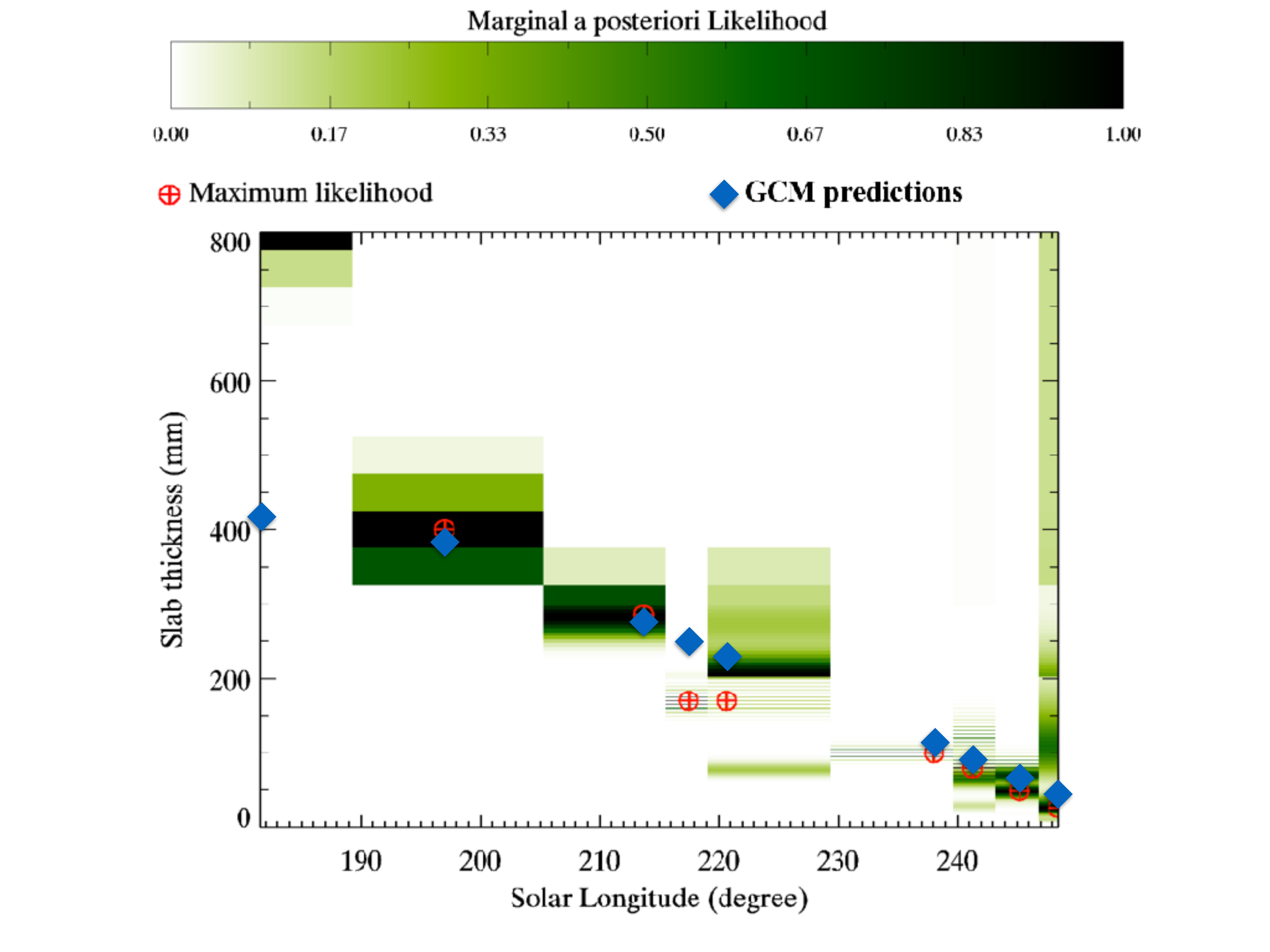}}\subfloat[\foreignlanguage{french}{Point CP2}]{\centering{}\includegraphics[width=0.5\textwidth,height=5.5cm]{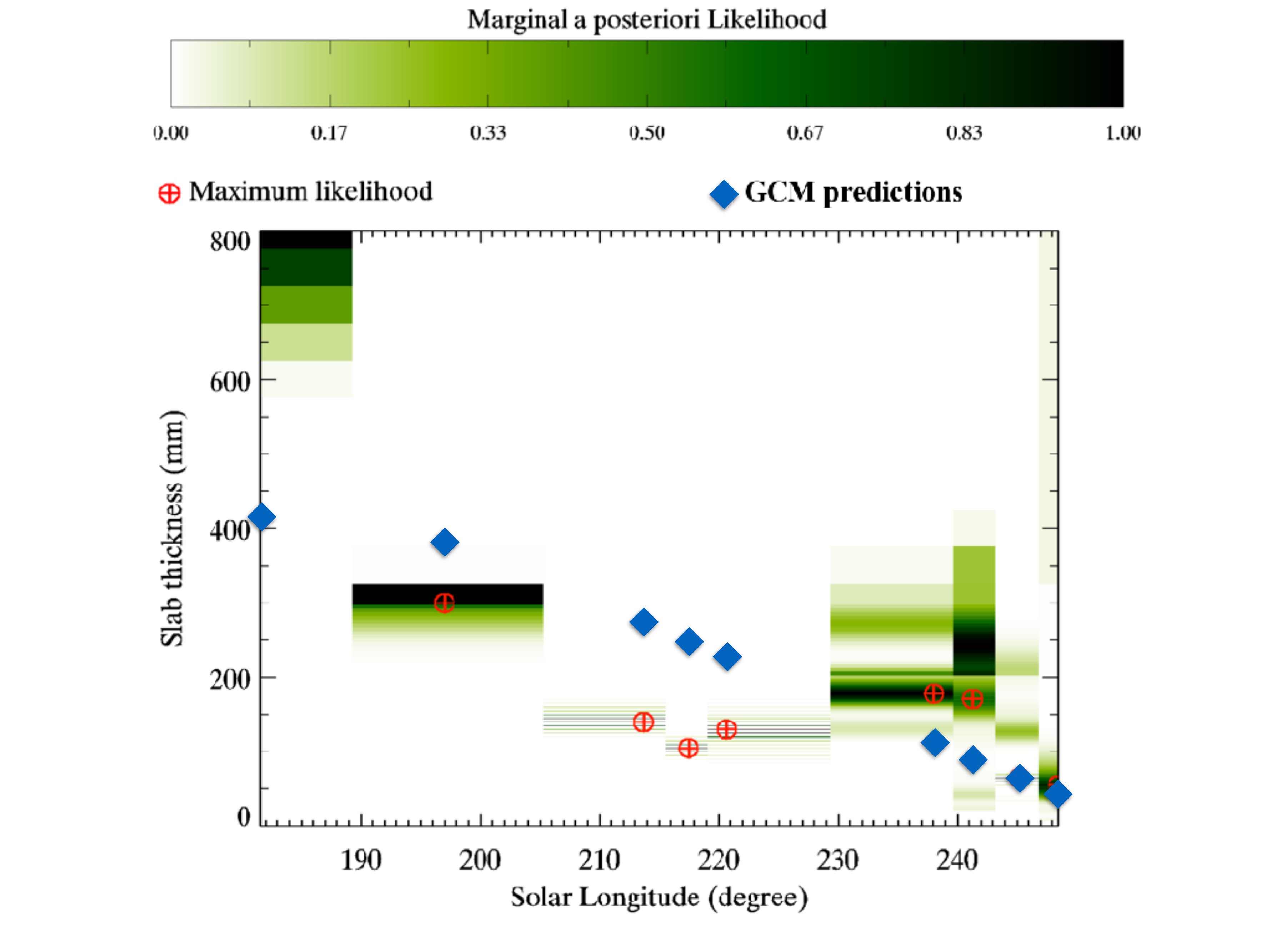}}\smallskip{}
\subfloat[\foreignlanguage{french}{Point DS1}]{\centering{}\includegraphics[width=0.5\textwidth,height=5.5cm]{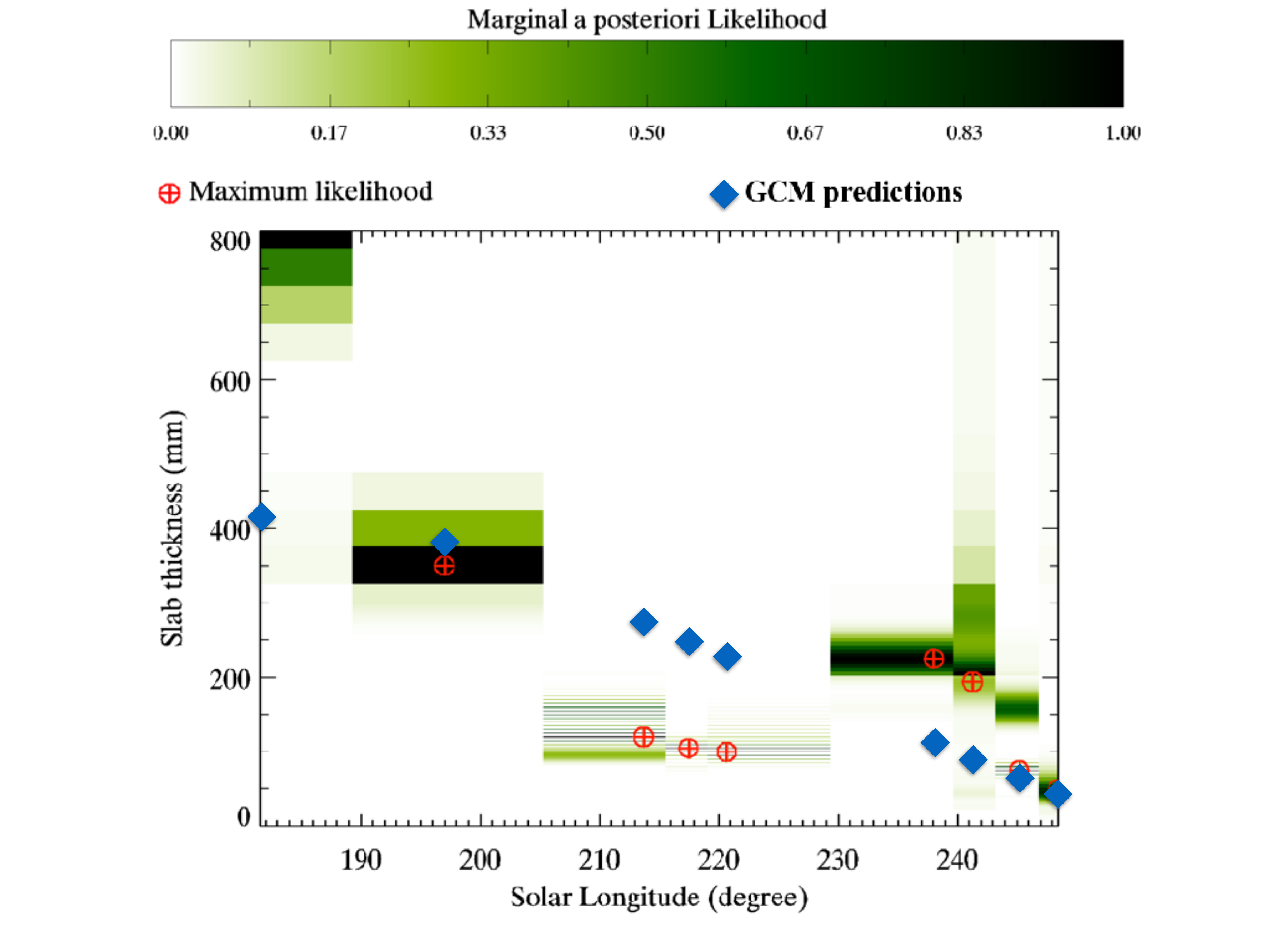}}\subfloat[\foreignlanguage{french}{Point DS2}]{\centering{}\includegraphics[width=0.5\textwidth,height=5.5cm]{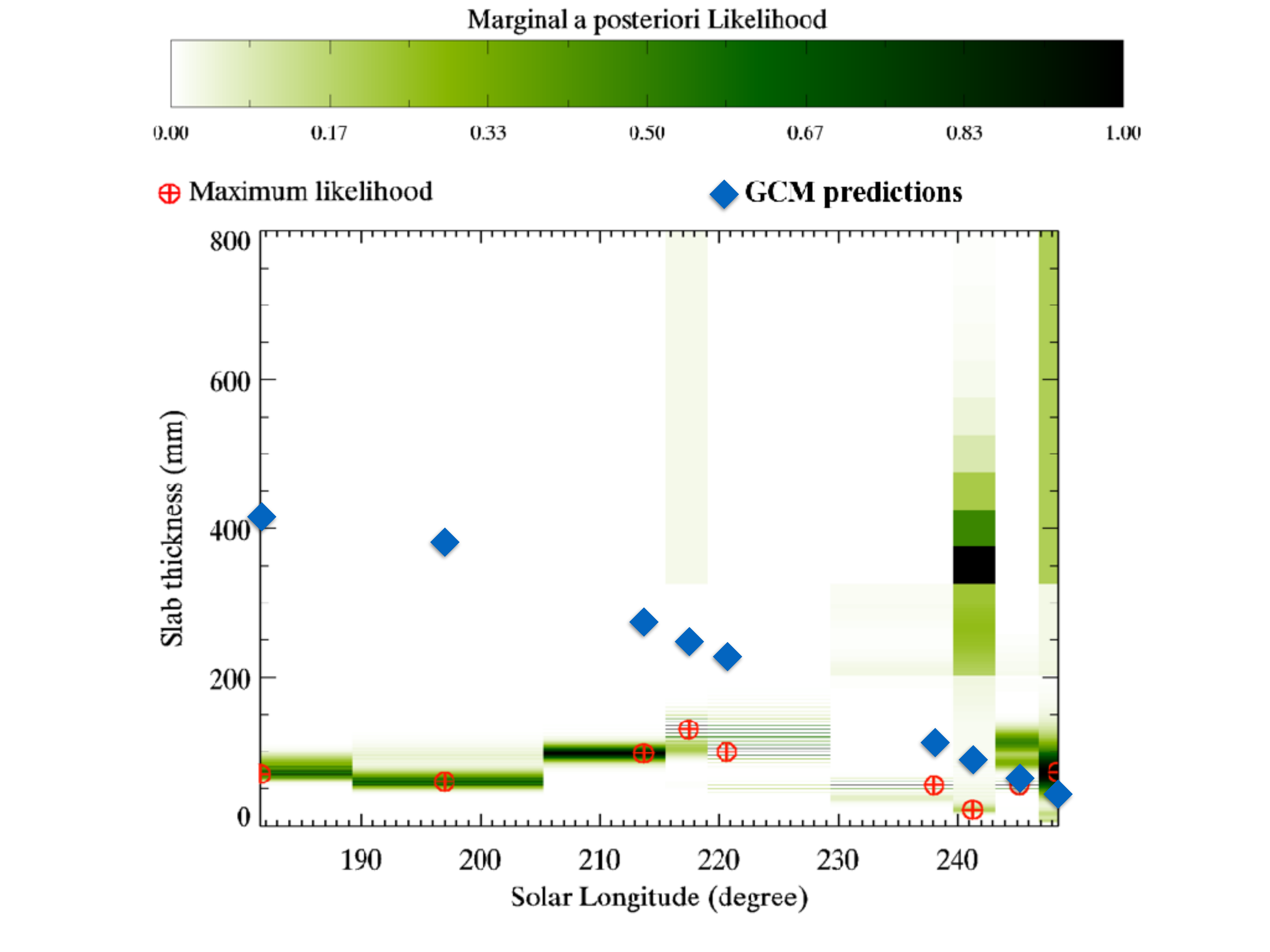}}\caption{\foreignlanguage{french}{\label{fig:Variations-h-control}\foreignlanguage{english}{Evolution
of the thickness of the top slab ice layer as a function of Ls for
(a) CP1 (b) CP2, (c) DS1 and (d) DS2. The shades of green represent
the values of the marginal a posteriori of the likelihood PDF, and
the red crosses point out the maximum value of the likelihood PDF
(best fit).}}}
\end{figure}

As a general rule, the thickness estimated by the inversion algorithm
for the ice layer decreases during spring, as is naturally expected.
Figure \ref{fig:Variations-h-control} presents the temporal ice thickness
for the four points of our study from early to late spring. They represent
on the same graph the posterior marginal probability density values
\LyXZeroWidthSpace \LyXZeroWidthSpace (eq. \ref{eq:Marginal_posterior_PDF-1})
for each inversion.

All points are following the expected sublimation process. One can
note a local variability of the thickness, on a scale that is not
taken into account by the GCM: the ice layer is thicker on the crests
of dunes (CP1) and thinner in the inter-dunes (CP2), which is expected.
Indeed, the inter-dunes show polygonal cracks (see Fig. \ref{fig:Context_map}b),
indicating possibly the presence of permafrost, and most probably
a water ice rich subsurface with a high thermal inertia. Conversely,
the dune is constituted of sand, with small grains with low thermal
inertia. A terrain with high thermal inertia will start to condense
CO$_{2}$ much later, and the dune will therefore start the spring
with a much thicker CO$_{2}$ ice layer than the inter-dunes. 

At the center of the dark spots, the thickness of the layer is lower
than that of the surrounding areas (the thicknesses at CP1 are compared
with those of DS1 and the thicknesses of CP2 with those of DS2), which
is consistent with the mechanism of cold jets proposed by \citealp{Piqueux2003}
and \citealp{Kieffer2006}: it is expected that the ice breaks where
it is most fragile, where the thickness of the layer is the lowest.
Alternatively, the points in the dark fans DS1 and DS2 could have
a lower slab thickness due to the enhanced sublimation due to darker
albedo because of the covering dust. In the dark spots of the inter-dune,
the thickness of the layer is low from the beginning of spring, which
is consistent with the activity of jets much earlier in these zones
(see fig. \ref{fig:Context_map}b), as seen by \citealp{Pommerol}
in the north. These are illuminated longer during the day than the
slope zones (the slopes are oriented towards the east or the west)
from the end of the polar night, and the greater roughness of the
inter-dune zones compared to the smooth zones of the slopes favor
a more important basal sublimation, and thus the activity is early
\citep{Pommerol}. It should be noted that the thickness remains almost
constant throughout the season, which is consistent with an effect
due to contamination by water ice proposed by \citep{Pommerol}, and
that will be discussed in section \ref{subsec:Contamination}. In
detail, DS1 and DS2 have a peculiar behavior that thickness is decreasing
from Ls=181.5 to Ls=210-220 and then increase to finally decrease.
We attribute this apparent thickening of the ice to the dust migration
within the ice, producing a optically thick layer of dust migrating
downwards.

\subsection{Contamination\label{subsec:Contamination}}

\subsubsection{Impurities proportions\label{subsec:Impurities}}

\begin{figure}
\centering{}\subfloat[\foreignlanguage{french}{CP1}]{\centering{}\includegraphics[width=0.5\textwidth,height=6cm]{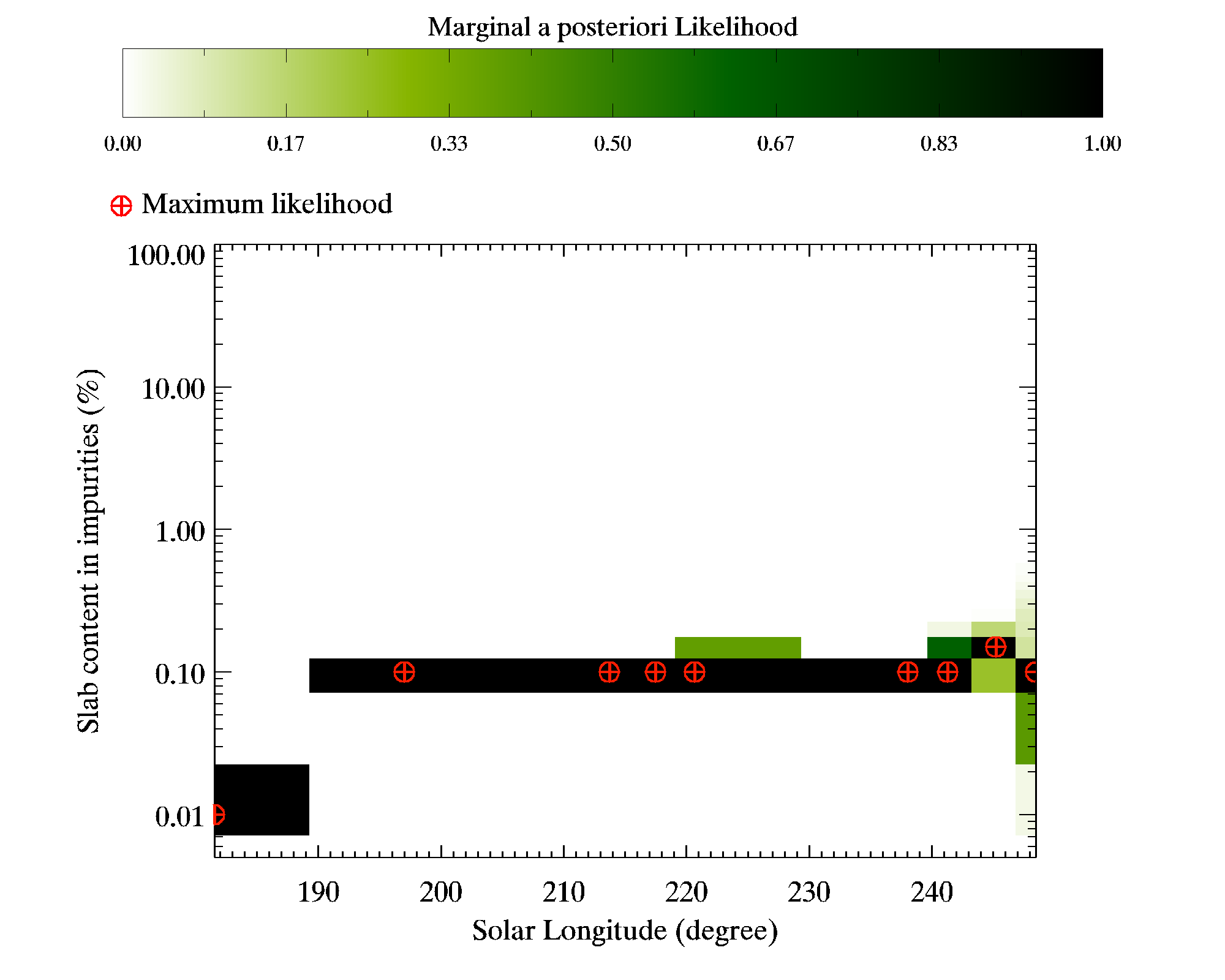}}\subfloat[\foreignlanguage{french}{CP2}]{\centering{}\includegraphics[width=0.5\textwidth,height=6cm]{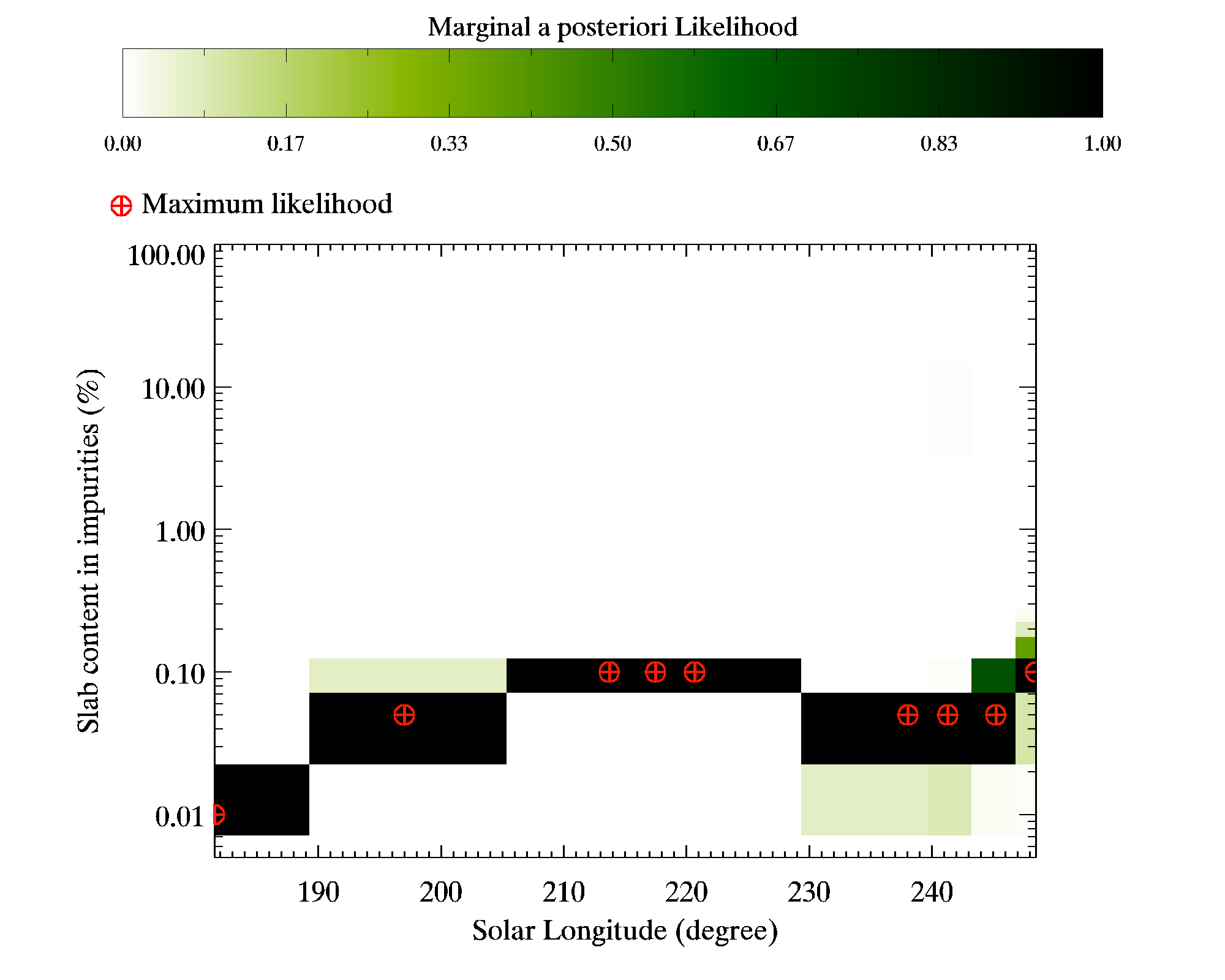}}\smallskip{}
\subfloat[\foreignlanguage{french}{DS1}]{\centering{}\includegraphics[width=0.5\textwidth,height=6cm]{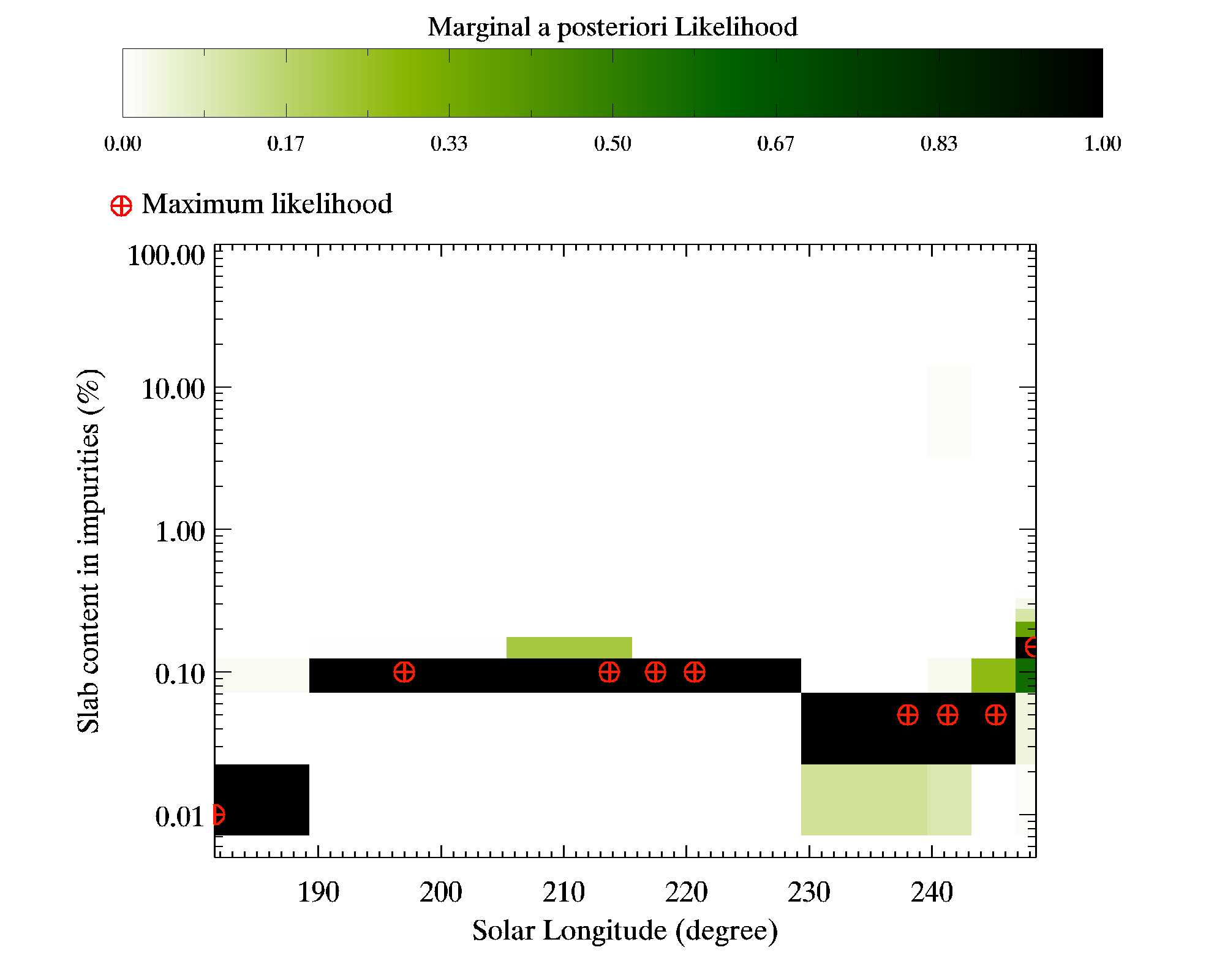}}\subfloat[\foreignlanguage{french}{DS2}]{\centering{}\includegraphics[width=0.5\textwidth,height=6cm]{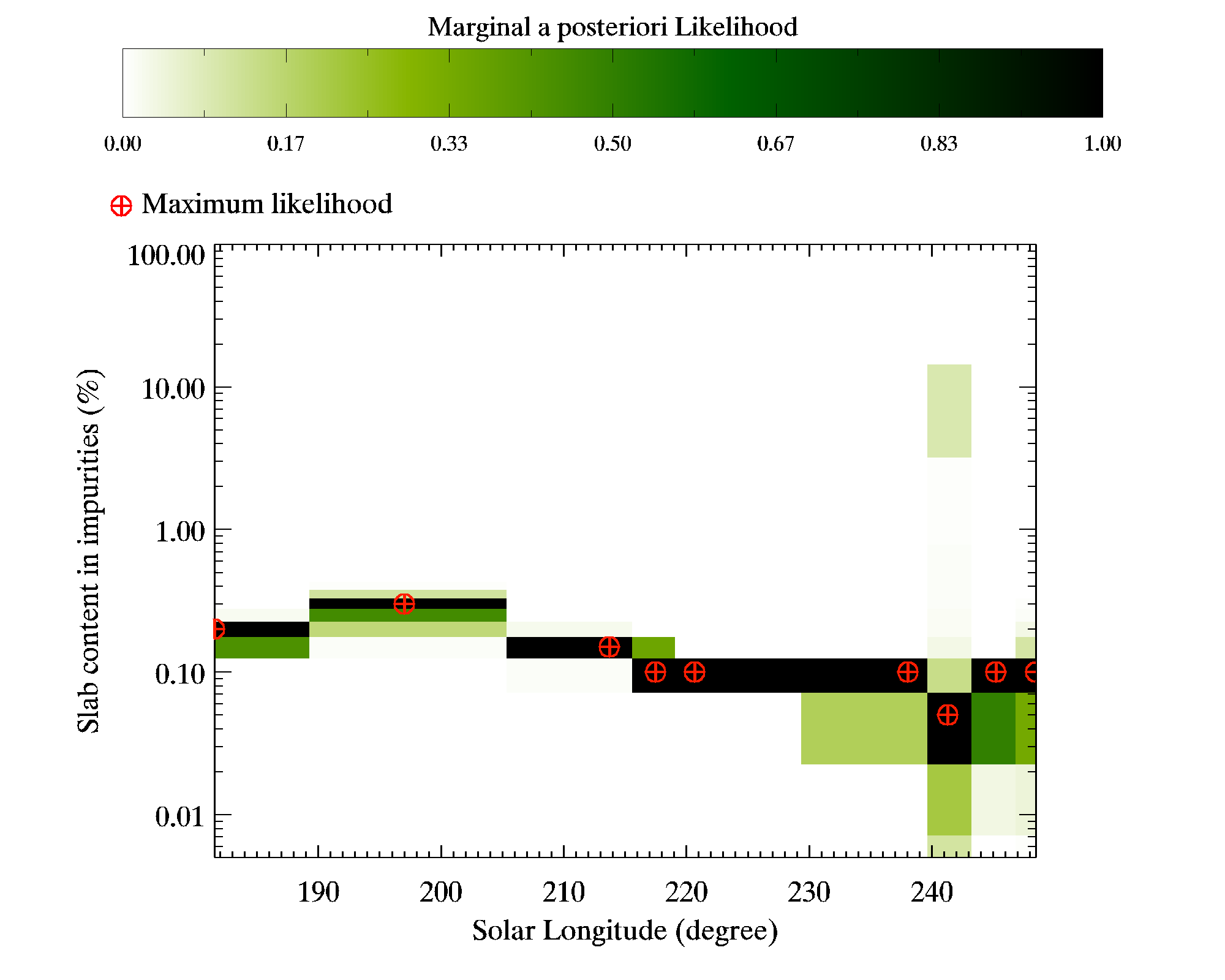}}\caption{\label{fig:Total-impuritites}Evolution of the total content in impurities
for (a) CP1 (b) CP2, (c) DS1 and (d) DS2. The shades of green and
the red crosses are interpreted in the same way as in the figure \ref{fig:Variations-h-control}.}
\end{figure}

\begin{figure}
\centering{}\subfloat[CP1]{\centering{}\includegraphics[width=8cm,height=6cm]{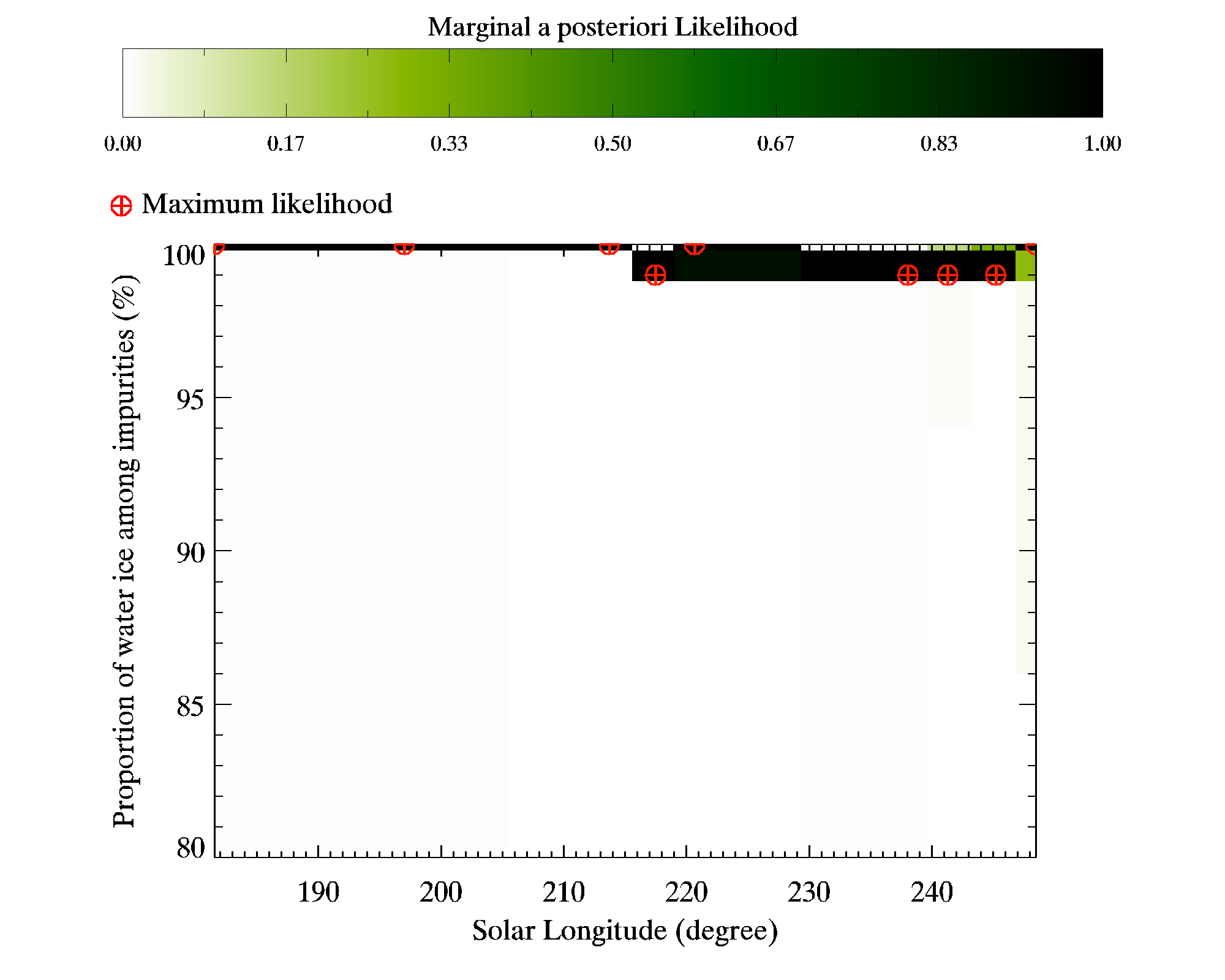}}\subfloat[\foreignlanguage{french}{CP2}]{\centering{}\includegraphics[width=8cm,height=6cm]{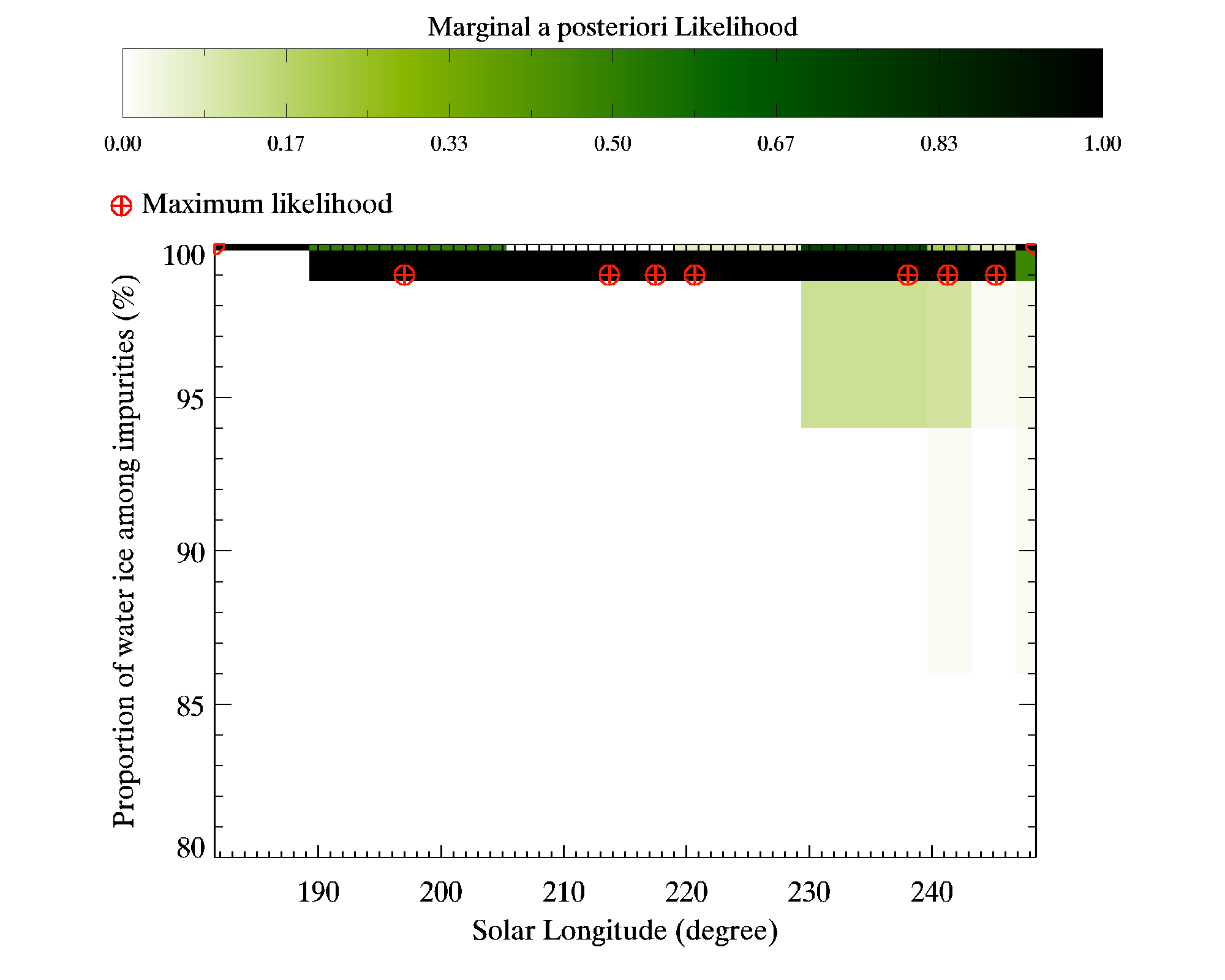}}\smallskip{}
\subfloat[\foreignlanguage{french}{DS1}]{\centering{}\includegraphics[width=8cm,height=6cm]{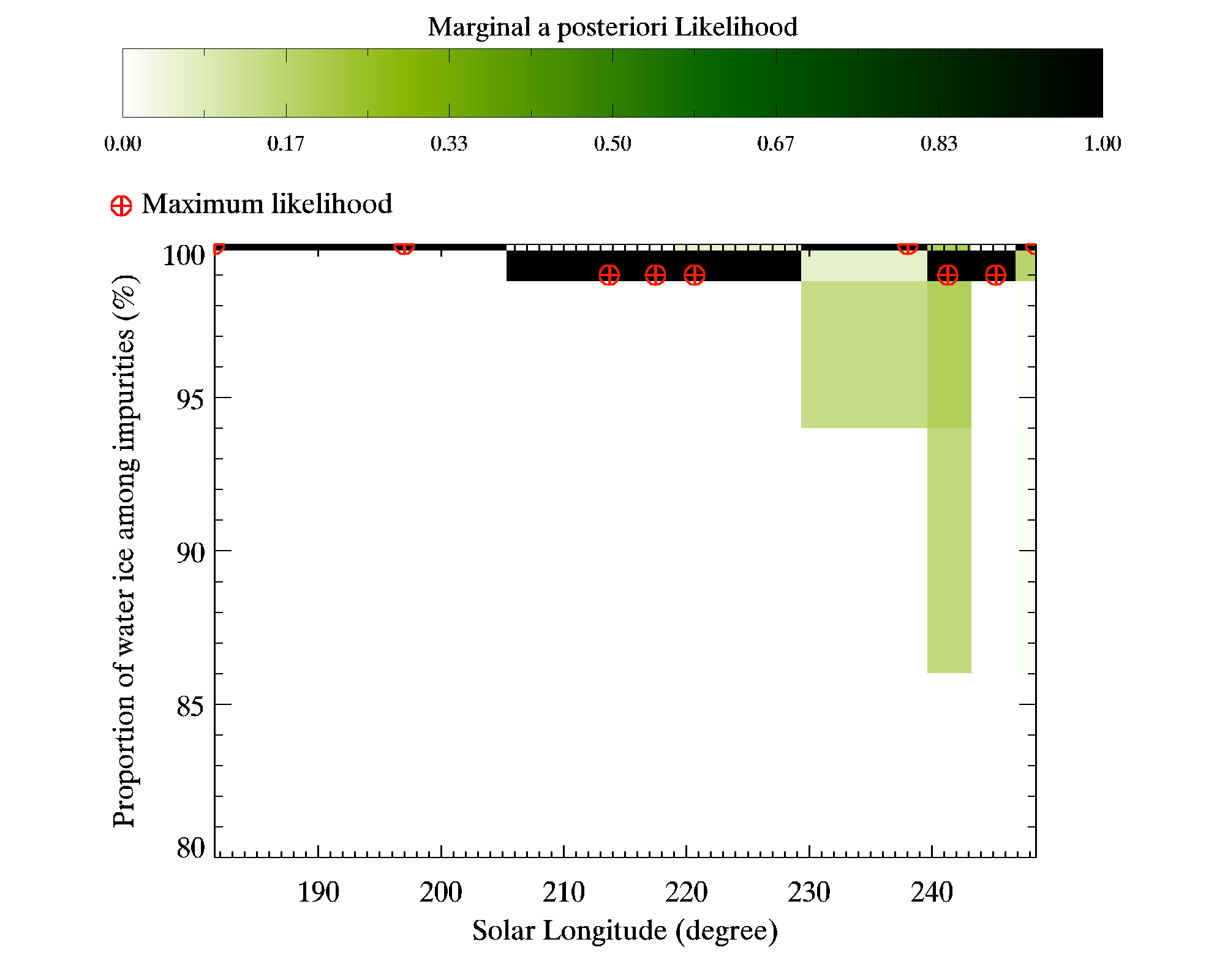}}\subfloat[\foreignlanguage{french}{DS2}]{\centering{}\includegraphics[width=8cm,height=6cm]{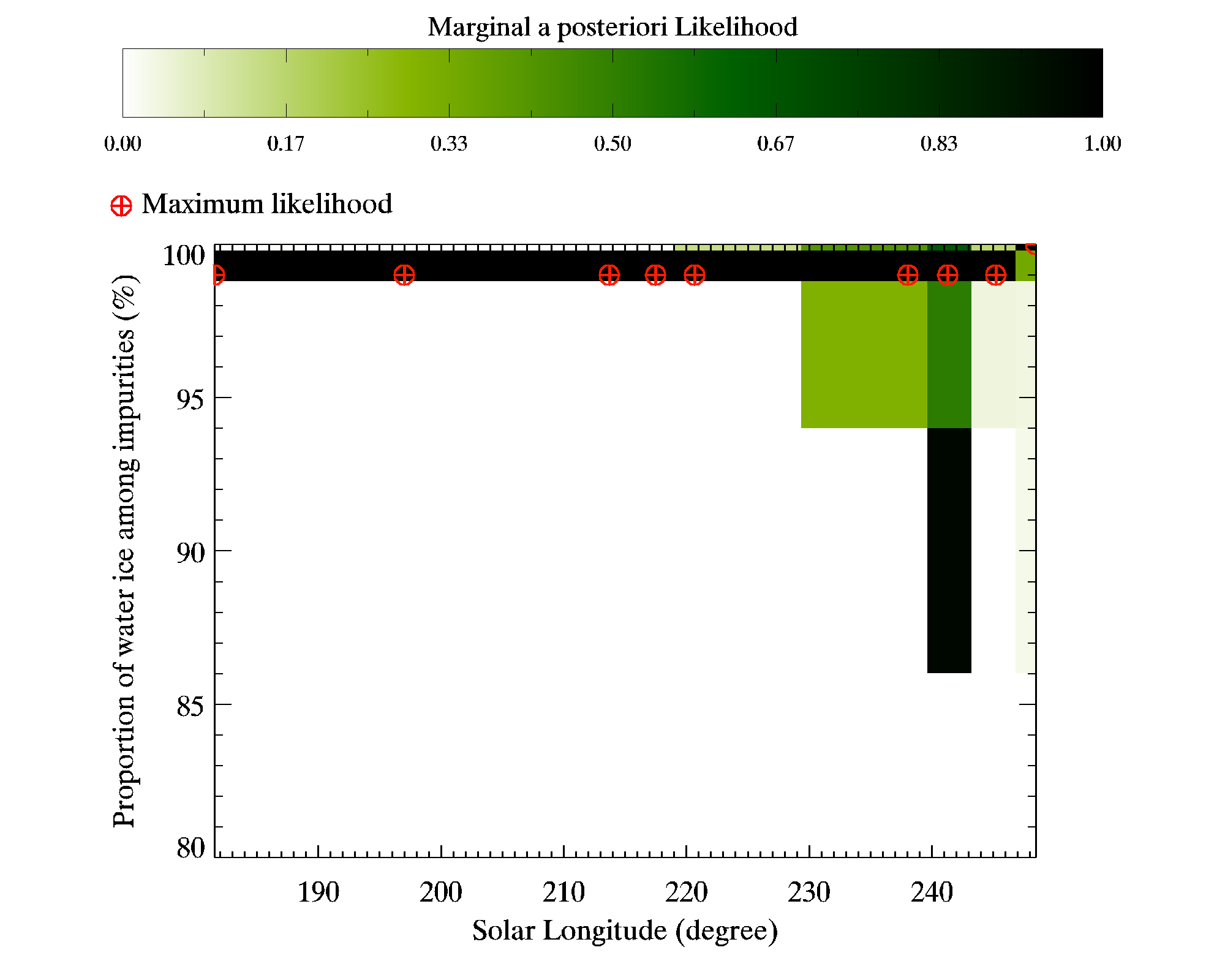}}\caption{\label{fig:WaterDustFraction-impuritites}Evolution of the proportion
of water ice among impurities (a) CP1 (b) CP2, (c) DS1 and (d) DS2.
The shades of green and the red crosses are interpreted in the same
way as in the figure \ref{fig:Variations-h-control}.}
\end{figure}

The total impurity content of the ice layer corresponds to the sum
of the volume proportions of water ice and dust in the matrix. As
can be seen in figure \ref{fig:Total-impuritites}, this content seems
to remain relatively stable between 0.01 \% and 0.1\% for the all
points. The total budget of impurities has to be investigated conjointly
with the respective proportions of dust and water ice among them,
represented on figure \ref{fig:WaterDustFraction-impuritites}. For
all the points studied, dust contamination seems very low. Indeed,
the relative proportions between water ice and dust estimated by the
inversion algorithm systematically give a proportion of ice water
greater than $99\,\mbox{\%}$, or a total content of dust always less
than $5\,\mbox{ppmv}$ (see Figure \ref{fig:WaterDustFraction-impuritites}).
This result tends to reinforce the hypothesis of a very effective
cleaning of dust \citep{kieffer2007cold,Portyankina2010}. On the
contrary, water ice is not cleaned, or at least much less efficiently
than dust because of a higher albedo. This behavior is expected, the
dust being low albedo (around 0.3) compared to water ice, which may
have a very high albedo, and the cleaning efficiency decreasing for
a growing albedo \citep{Portyankina2010}. The graphs representing
the total impurity content (fig. \ref{fig:Total-impuritites}) can
therefore be directly interpreted as a volumetric water content (with
an error of less than $1\,\mbox{\%}$).

In details, the two points studied in the dark spots show different
behaviors: the DS1 point appears to have a level of contamination
similar to that of the control points. Conversely, the DS2 point in
the inter-dune appears to be considerably more contaminated with water
ice than all other points studied, displaying impurity contents up
to 5 times higher than other locations in early spring. The DS2 contamination
then decreases in a few weeks to stabilize at the same level as the
surrounding points.

The behavior of DS2 is consistent with the jet mechanism. Indeed,
the triggering of a jet of gas is supposed to be a violent event,
which tears material from the regolith and deposits it at the top
of the layer of ice. It is therefore expected that there will be more
contamination in the vicinity of dark spots. Conversely, DS1's behavior
is unexpected: its level of contamination is equivalent to that of
the environmental control point. The apparent inconsistency with the
jet mechanism can be resolved if most of the material ejected by the
cold jet do not stay in the vicinity of the jet, but is either carried
away by the wind, roll down the slope or is efficiently cleaned. It
would also mean that the darkness of spot around the jet is a photometric
effect. In addition, the upward sublimation flux of increasing intensity
during spring favors a transport of dust and water ice grains down
from the ridge of the dune to the inter-dunes, by decreasing the friction
coefficient between the grains and the slope. This would explain the
formation of flows from dark spots, such as the one in Fig. \ref{fig:hirise_series}
from $L_{S}=213\text{\textdegree}$. In this case, it would be expected
to find greater contamination in the inter-dune, or at least at the
foot of the slopes. The follow-up of a control point precisely at
the bottom of the slope would make it possible to decide on this point.
Another possible explanation is that the jets in the inter-dunes contain
much more water ice than the jets in the ridges: as the inter-dunes
subsurface is most probably water ice rich (polygons) and the jet
mechanism may pull water out from the subsurface to the fan \citep{Kereszturi2011}.
It will later enrich the layer in water ice. Conversely, the dune
is constituted of sand, and the jet will only pull dust, that will
then be efficiently cleaned before the next measurement \citep{Pommerol2011}. 

\subsubsection{Water grain size\label{subsec:Water-grain-size}}

\begin{figure}
\centering{}\subfloat[\foreignlanguage{french}{CP1}]{\centering{}\includegraphics[width=8cm,height=6cm]{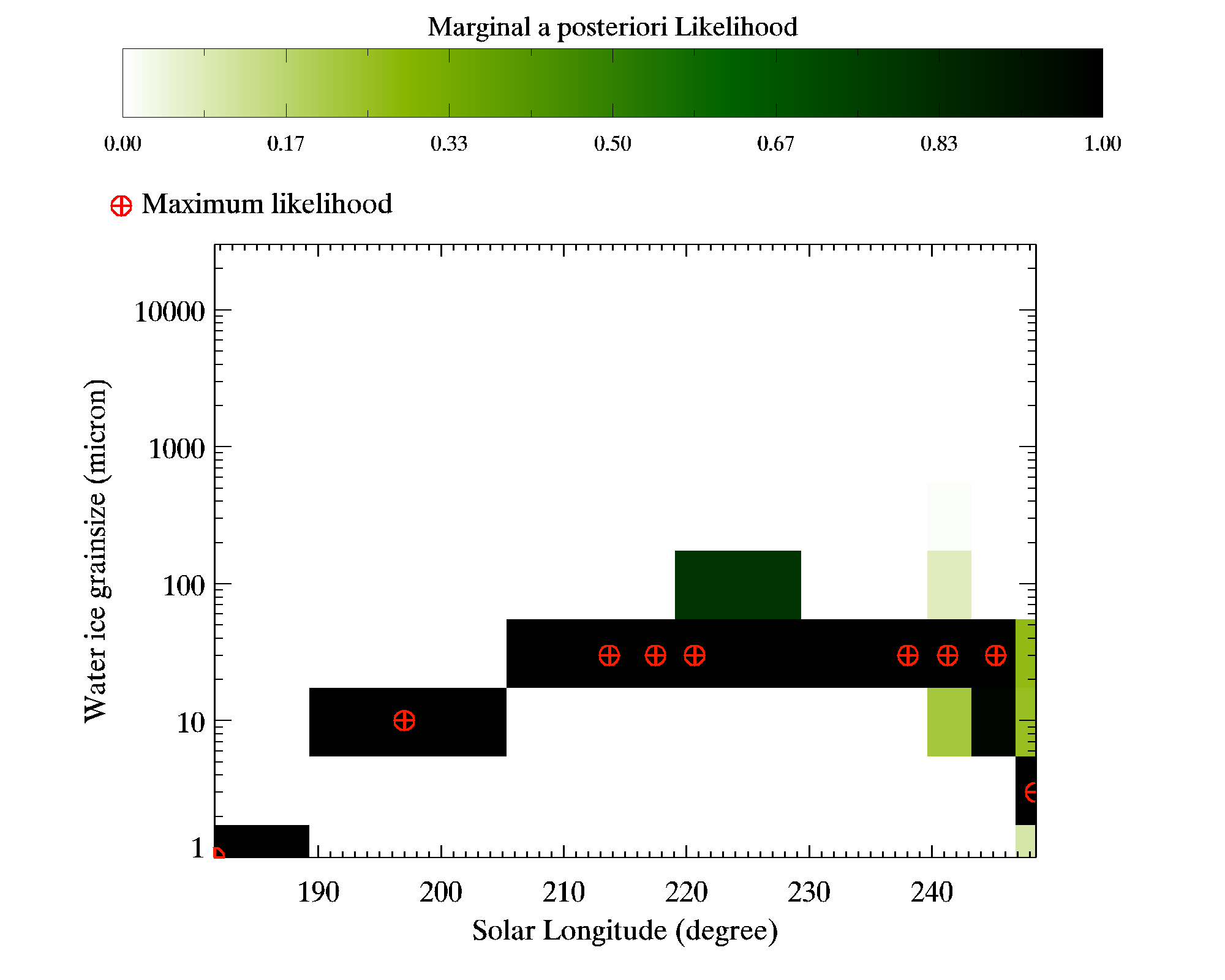}}\subfloat[\foreignlanguage{french}{CP2}]{\centering{}\includegraphics[width=8cm,height=6cm]{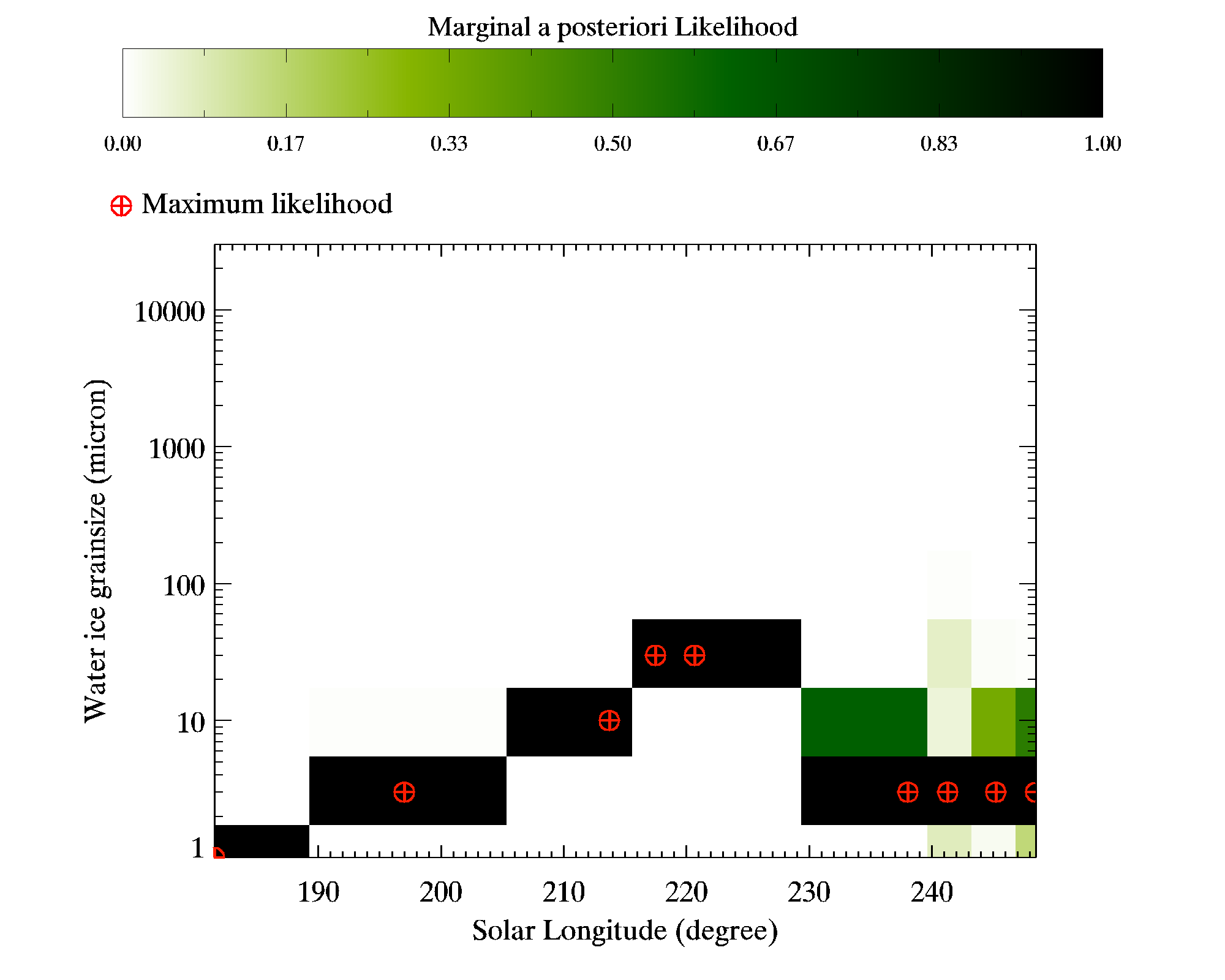}}\smallskip{}
\subfloat[\foreignlanguage{french}{DS1}]{\centering{}\includegraphics[width=8cm,height=6cm]{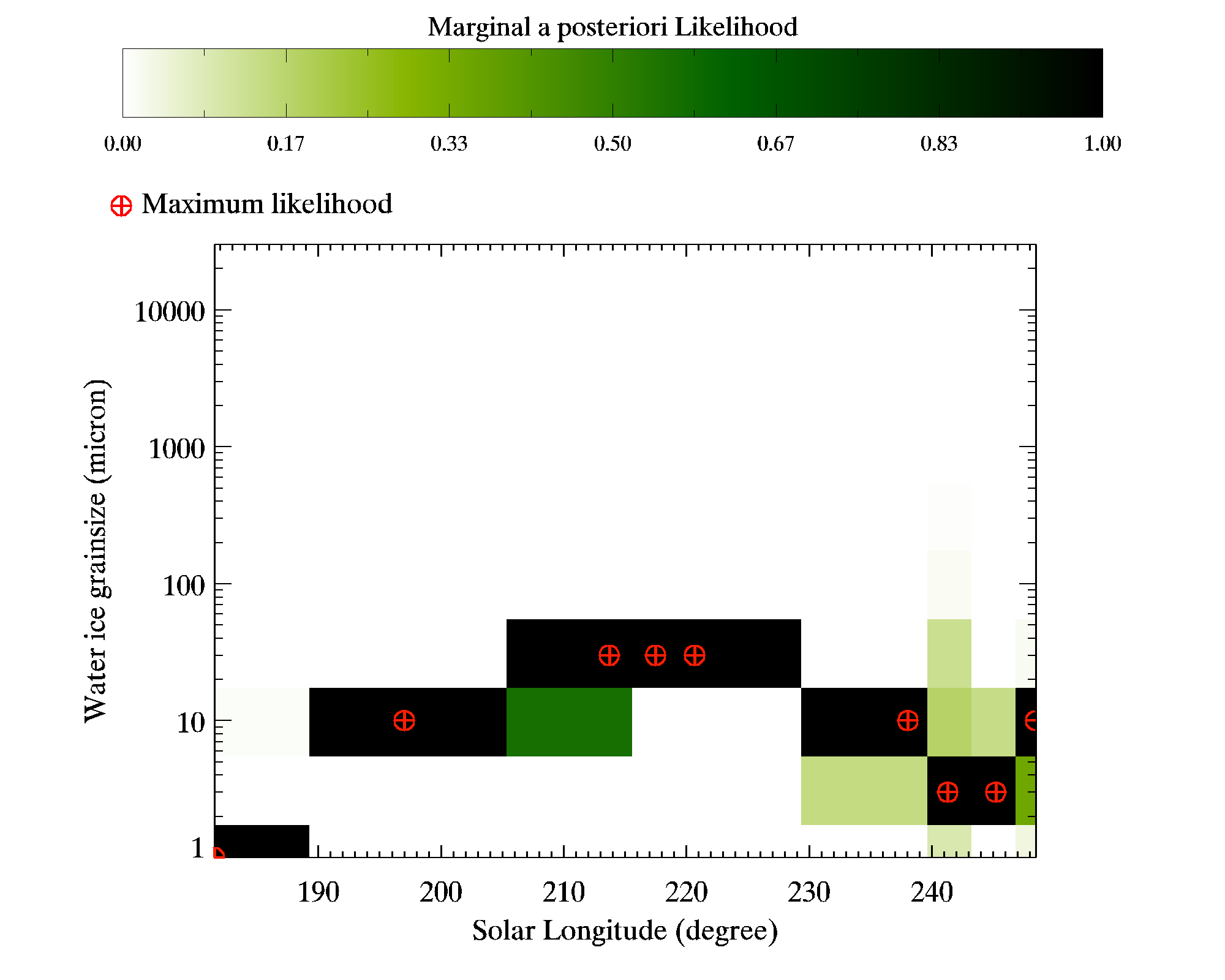}}\subfloat[\foreignlanguage{french}{DS2}]{\centering{}\includegraphics[width=8cm,height=6cm]{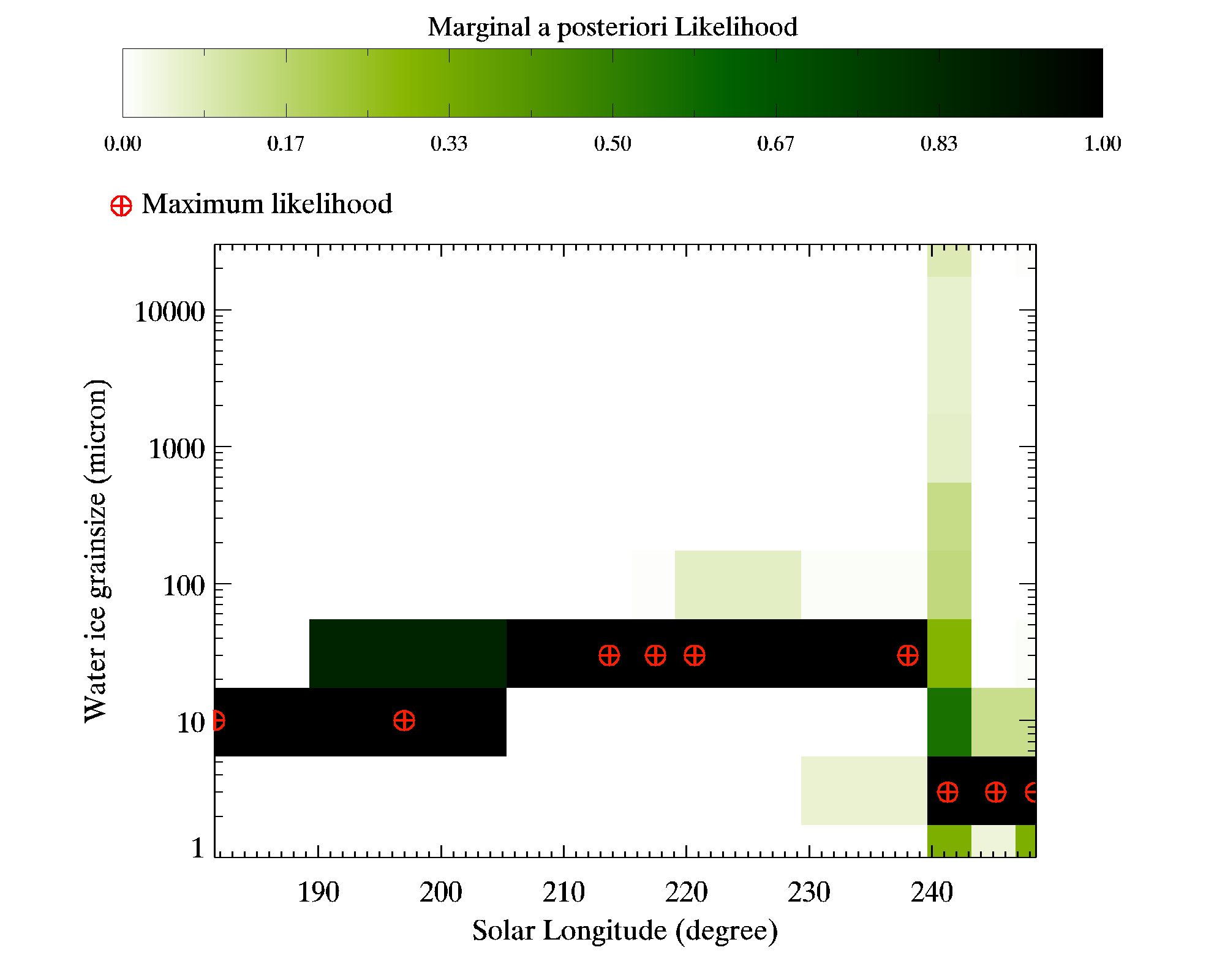}}\caption{\label{fig:WaterIceGrainsize}Evolution of the water ice grain diameter
for (a) CP1 (b) CP2, (c) DS1 and (d) DS2. The shades of green and
the red crosses are interpreted in the same way as in the figure \ref{fig:Variations-h-control}.}
\end{figure}

The figure \ref{fig:WaterIceGrainsize} shows that the grain size
increases during the spring for all the studied points, up to about
$L_{S}=238\text{\textdegree}$ from 1 microns to 50 microns. Then
grain size suddenly decreases up to few microns at the end of the
season. Variations in water ice grain size are not expected during
the sublimation of the seasonal CO$_{2}$ deposits. Indeed, water
is supposed to be thermalized by CO$_{2}$ ice, as carbon dioxide
has a condensation point much lower than water. So an evolution in
the grain size means that water ice may have a significant mobility,
even at the very beginning of the spring. This is unexpected and a
mechanism able to explain the variations both in grain size and total
content will be detailed in the following subsections. 

\subsubsection{Water escape\label{subsec:Water-escape}}

Water ice at the surface is in contact with CO$_{2}$ ice throughout
the spring, which is a cold trap. Therefore H$_{2}$O is not expected
to sublimate as long as the ice of CO$_{2}$ has not completely disappeared.
To remain the total water budget constant in a shrinking CO$_{2}$
slab, we expect an increase of volumetric water impurity. However,
the combined analysis of the previous subsections clearly show that
this is not the case. So there is an water escape process. This escape
can be explained by the spring activity of jets. Indeed, during an
ejection, part of the water ice entrained by the jet can remain suspended
in the atmosphere, carried by the ambient sublimation flux \citep{Kieffer2000}.
In this case, the fine part of the water ice grains must gradually
be eliminated over repeated ejections, and the grain size must therefore
increase during the spring, as observed (fig. \ref{fig:WaterIceGrainsize}). 

Note that water ice remain throughout spring present in amounts that
are far lower than CO$_{2}$ (about $0.1\%$), and that there is no
clear albedo variations such as those described by \citealp{Pommerol2011}.
The formation of a thin water frost layer at the top of the subliming
CO$_{2}$ ice layer would create an albedo increase that is not seen
here. Therefore it is supposed that water ice is always in intimate
mixing with CO$_{2}$ ice. Due to removal from the bottom and deposition
at the top, there still may be a vertical variation of the water ice
content inside the CO$_{2}$ ice layer, that is not taken into account
in the inversion. All quantities are vertically averaged. 

\subsubsection{Dust in linear mix\label{subsec:Poussi=0000E8res}}

\begin{figure}
\centering{}\subfloat[\foreignlanguage{french}{Point CP1}]{\centering{}\includegraphics[width=8cm,height=6cm]{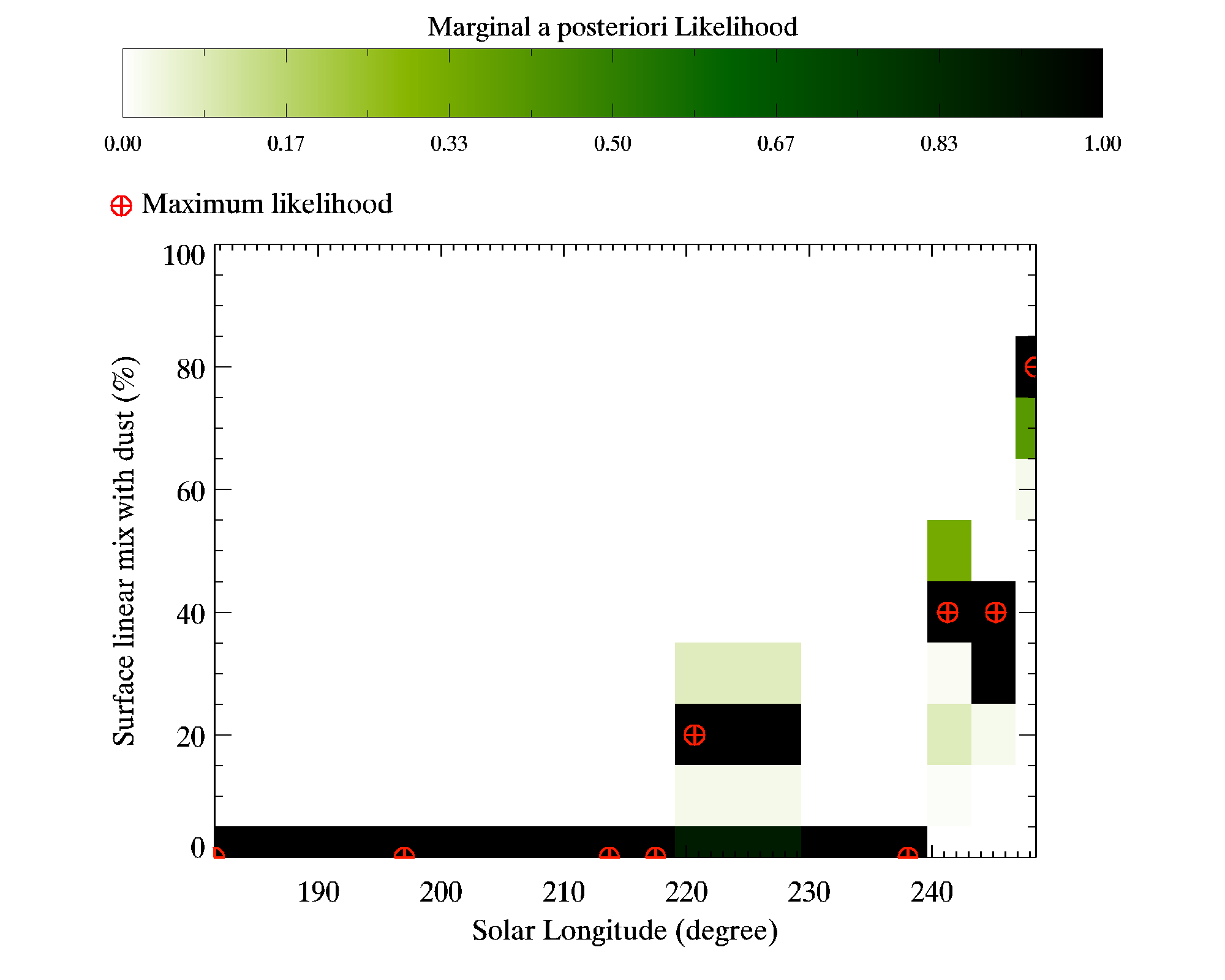}}\subfloat[\foreignlanguage{french}{Point CP2}]{\centering{}\includegraphics[width=8cm,height=6cm]{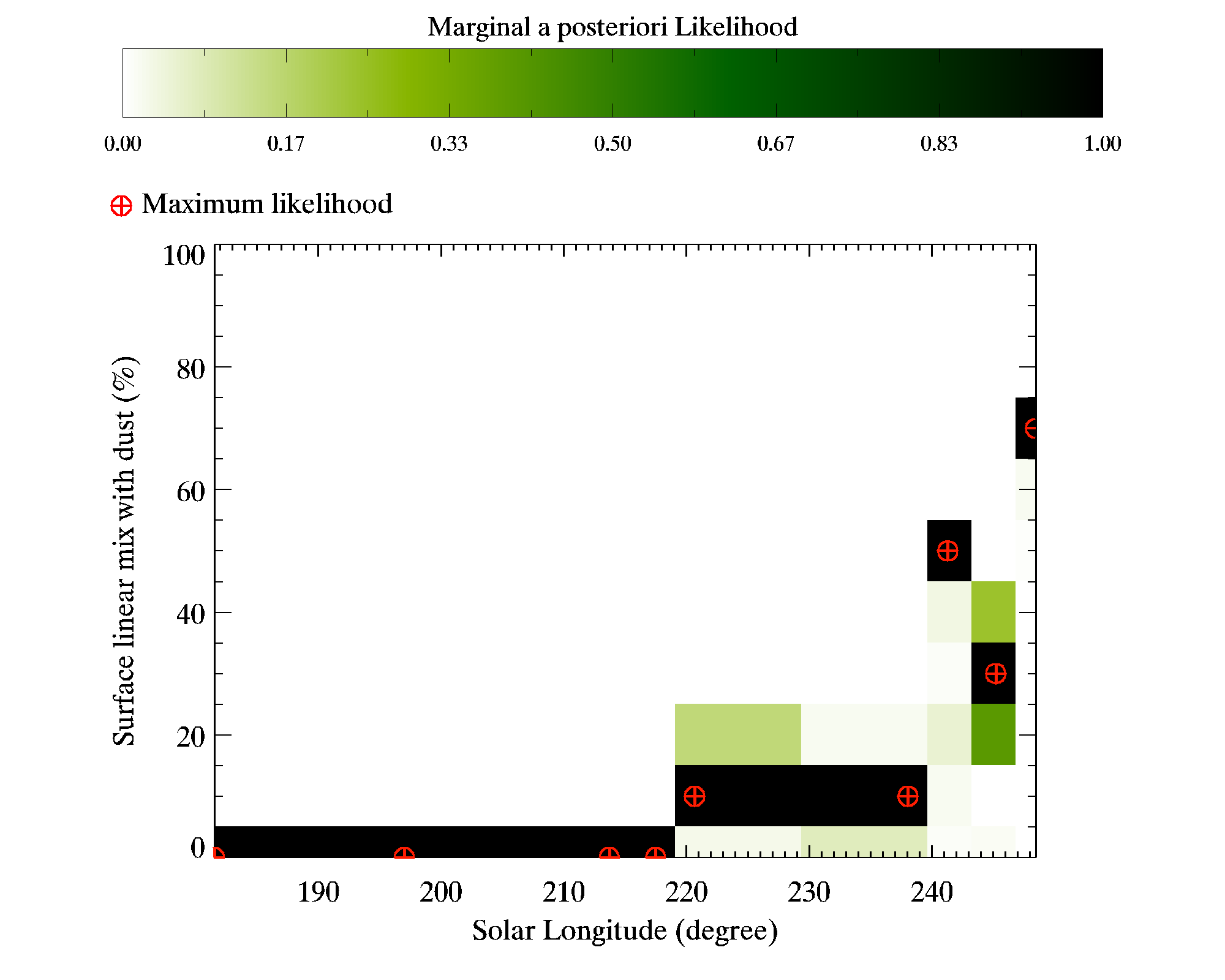}}\smallskip{}
\subfloat[\foreignlanguage{french}{Point DS1}]{\centering{}\includegraphics[width=8cm,height=6cm]{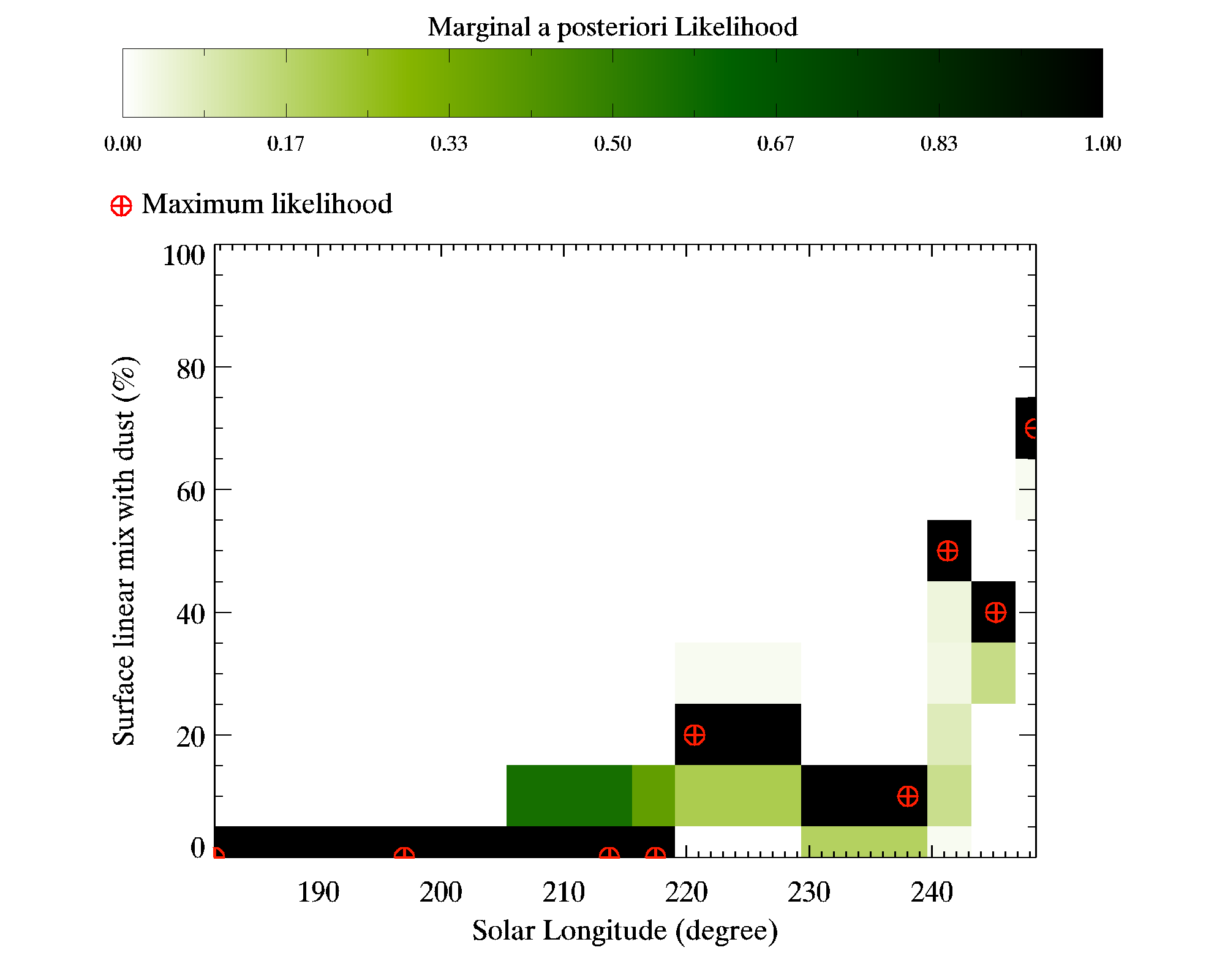}}\subfloat[\foreignlanguage{french}{Point DS2}]{\centering{}\includegraphics[width=8cm,height=6cm]{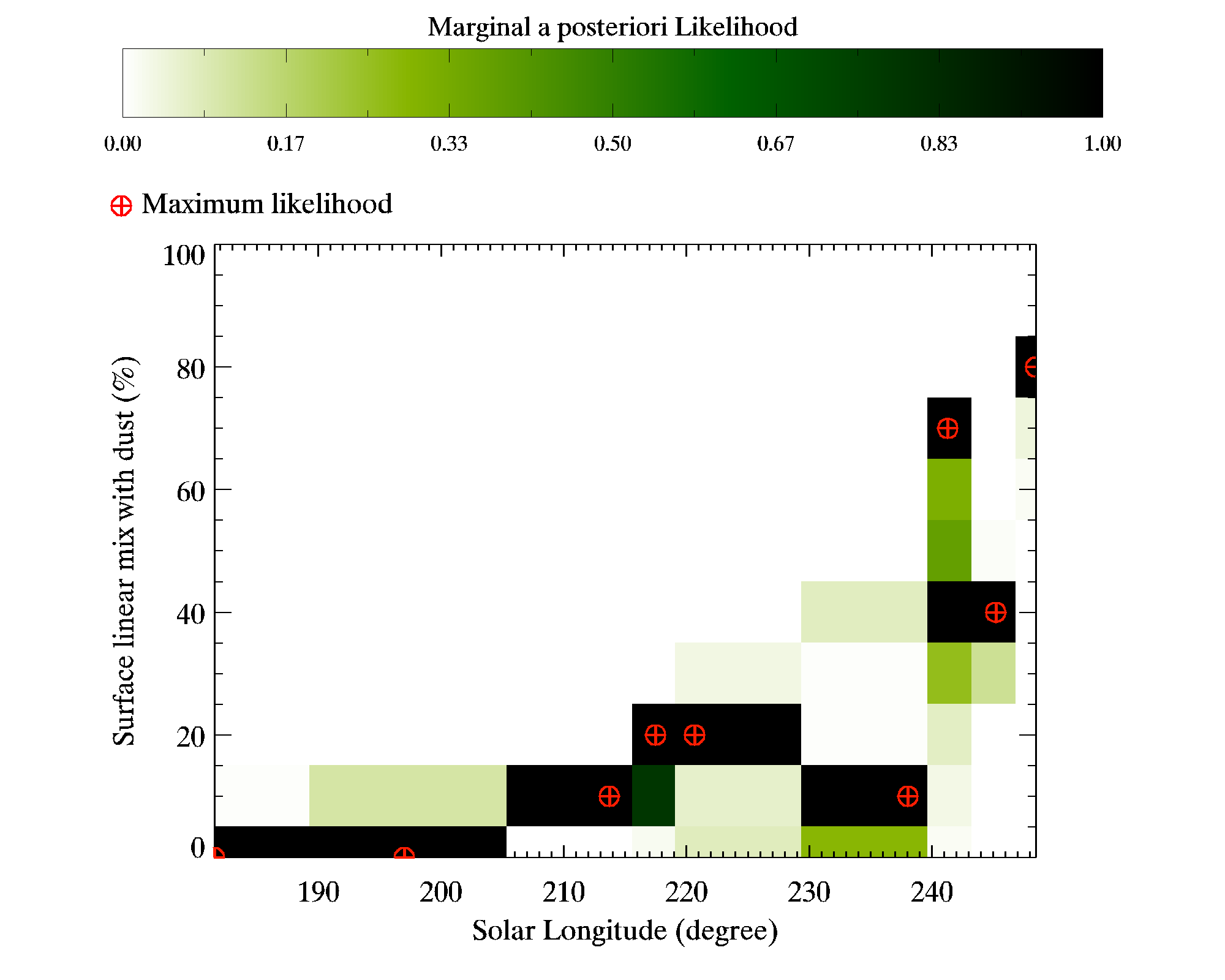}}\caption{\label{fig:LinearDustMix}Evolution of the proportion of dust in linear
mix for (a) CP1 (b) CP2, (c) DS1 and (d) DS2. The shades of green
and the red crosses are interpreted in the same way as in the figure
\ref{fig:Variations-h-control}.}
\end{figure}

The jet mechanism proposed to explain dark spots involves the deposition
of a layer of dust and possibly water ice at the top of the layer
after ejection. If the layer is thick enough locally, one should observe,
either only dust in the center of the dark spot, or a linear mixture
between two types of surface: dust and ice. We can see in Fig. \ref{fig:LinearDustMix}
that there is no significant difference in terms of sub-pixel geographic
mixing of dust and ice between points. Moreover, it can be noted that
all the points studied exhibit a similar behavior, with a small peak
of contamination around $L_{S}=220\text{\textdegree}$ that can be
explained by a possible dust storm, as also see in the atmospheric
dust content estimate (see table \ref{tab:R=0000E9capitulatif-des-observations}).
Then, the surface proportion of dust increase from $L_{S}=238\text{\textdegree}$
to the end. This is probably due to the fact that from this date the
CO$_{2}$ has been totally sublimated on some areas of the image.
This explanation is consistent with the sudden decrease in grain size
observed in fig. \ref{fig:WaterIceGrainsize}. Indeed, CO$_{2}$ ice
forces the temperature of the media in contact with it to its equilibrium
temperature (about $150\,\mbox{K}$), but when the CO$_{2}$ has disappeared,
dust and water ice are no longer buffered by this CO$_{2}$ condensation
temperature. In this case, the temperature increases, and the water
ice finally sublimates. However, there are still ice-covered areas
of CO$_{2}$ a few meters away, and water could be trapped again.
Water re condensed at the surface, forming a fine frost, explaining
the sudden decrease in the size of the observed grains. This mechanism
was also observed in the north, the sublimation of seasonal deposits
leading to the formation of a ring of water frost following the recession
of the seasonal cap \citep{Wagstaff2008,Appere2011,appere2012cycle,Pommerol}.

It is worth to remind that, on the wavelength range $1\,\mbox{\ensuremath{\mu} m}-2.6\,\mbox{\ensuremath{\mu} m}$,
the pure dust has a spectral behavior very monotonic, quasi-linear.
In fact, it is difficult to distinguish the effects of aerosols in
the atmosphere from dust within the ice layer. It is therefore expected
to have high uncertainty on dust contamination estimates by the inversion
algorithm. Actually, this is not the case in our results: the amounts
of dust are very well constrained, even tacking into account the uncertainties
due to atmospheric correction. We attribute this behavior to the non-linear
coupling between CO$_{2}$ bands and dust within the slab. The same
behavior may also occur in case of differentiation between linear
surface mixing with non-linear impurities. Again our results show
that there is enough information in the spectral data to decipher
this situation, most probably due to the non-linearities.

It would still be interesting to test the algorithm with other atmospheric
correction methods. For example, the photometric properties of aerosols
can be used to better isolate them and improve the correction \citep{Doute2013,Doute2014}.

\section{Discussion}

\begin{figure}
\begin{centering}
\subfloat[]{\centering{}\includegraphics[width=0.33\textwidth]{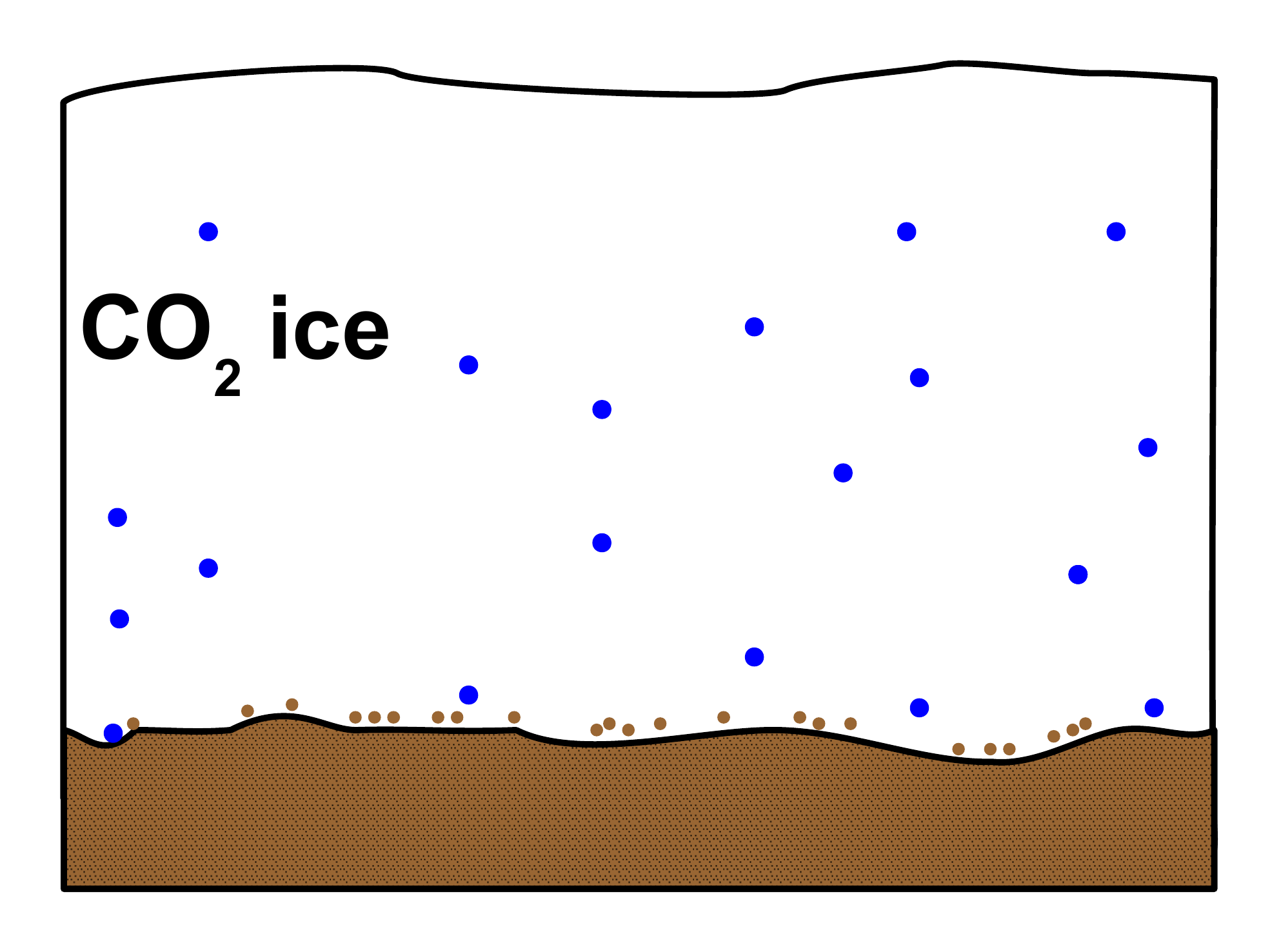}}\subfloat[]{\centering{}\includegraphics[width=0.33\textwidth]{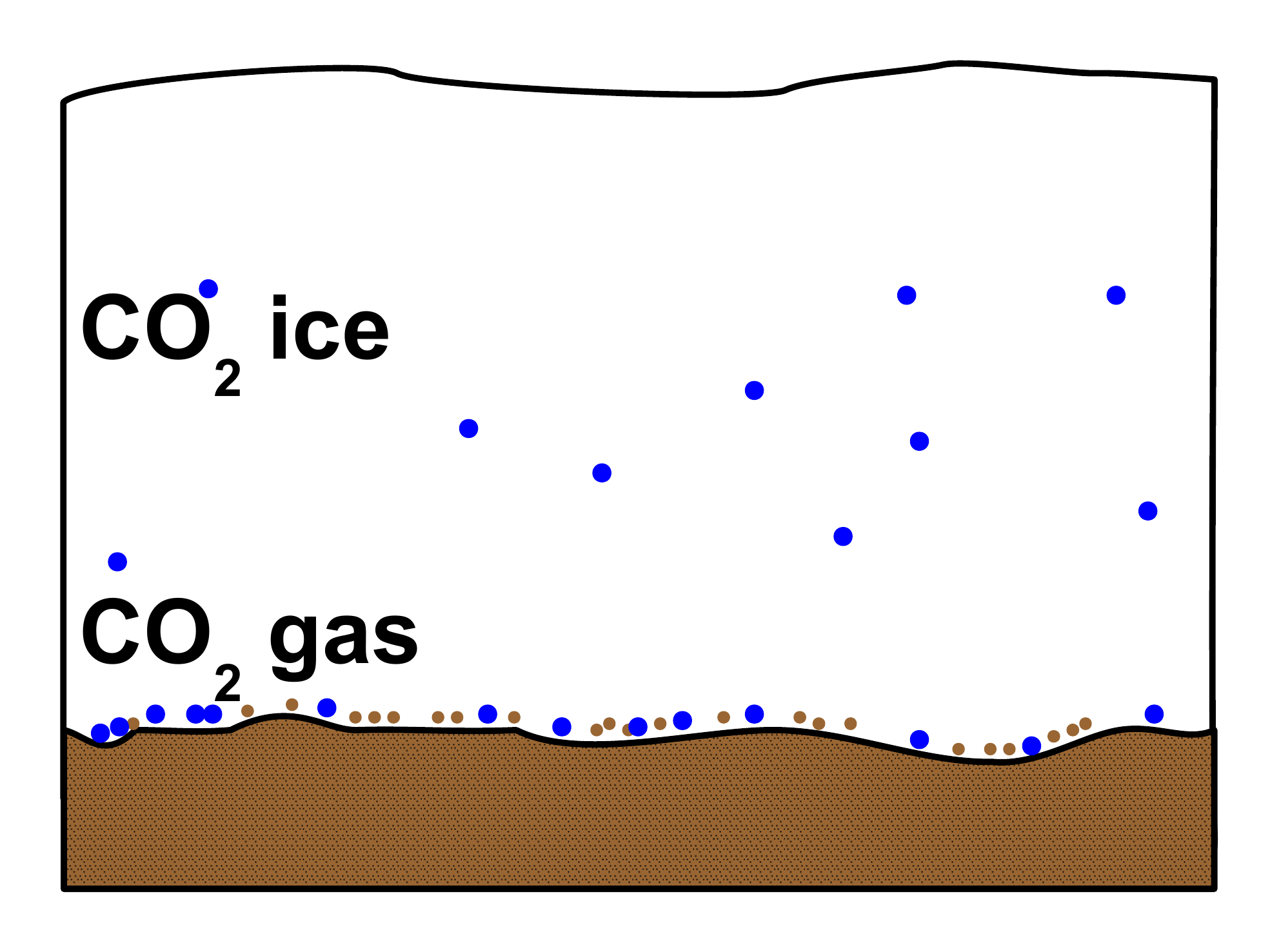}}\\
\subfloat[]{\centering{}\includegraphics[width=0.33\textwidth]{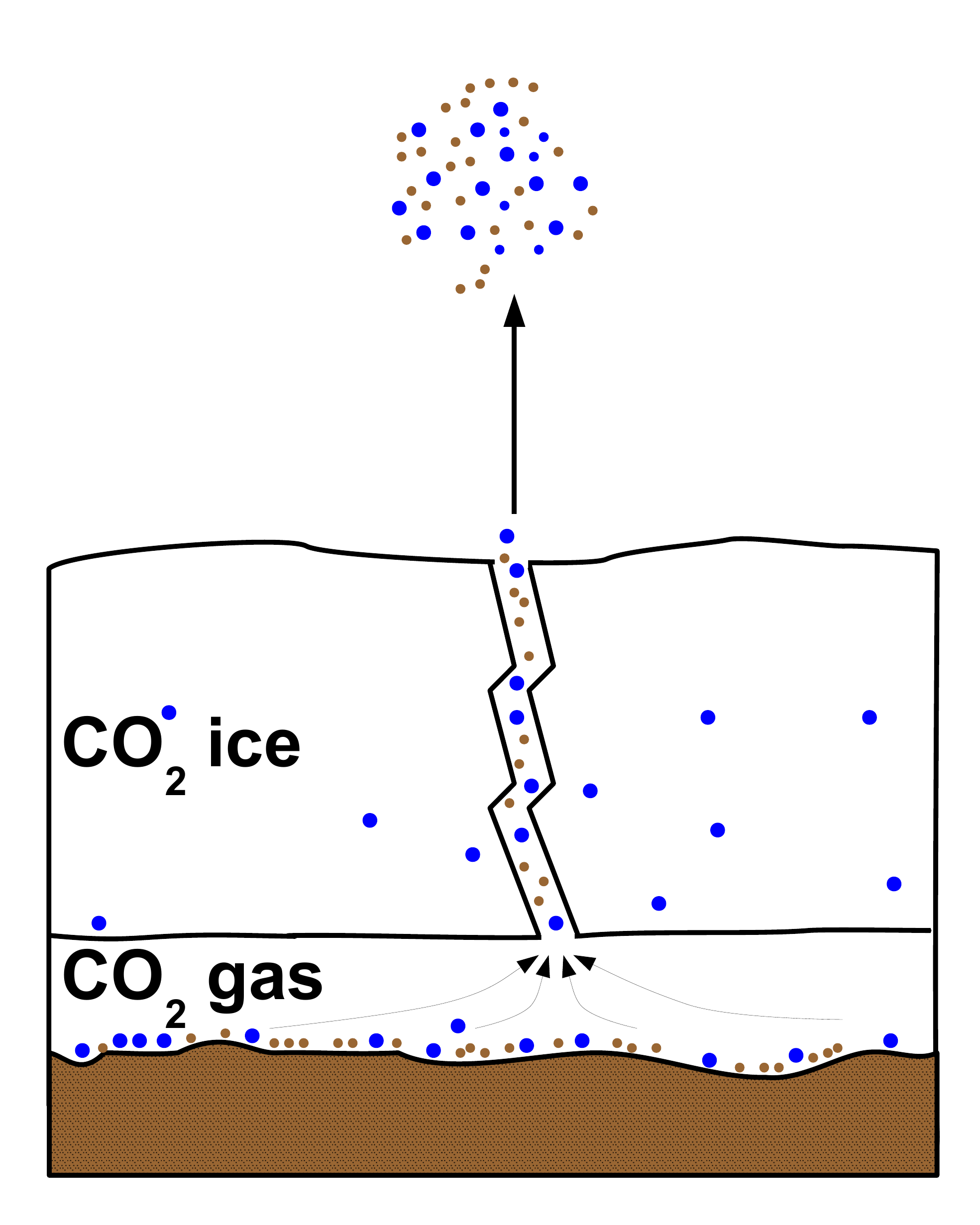}}\subfloat[]{\centering{}\includegraphics[width=0.33\textwidth]{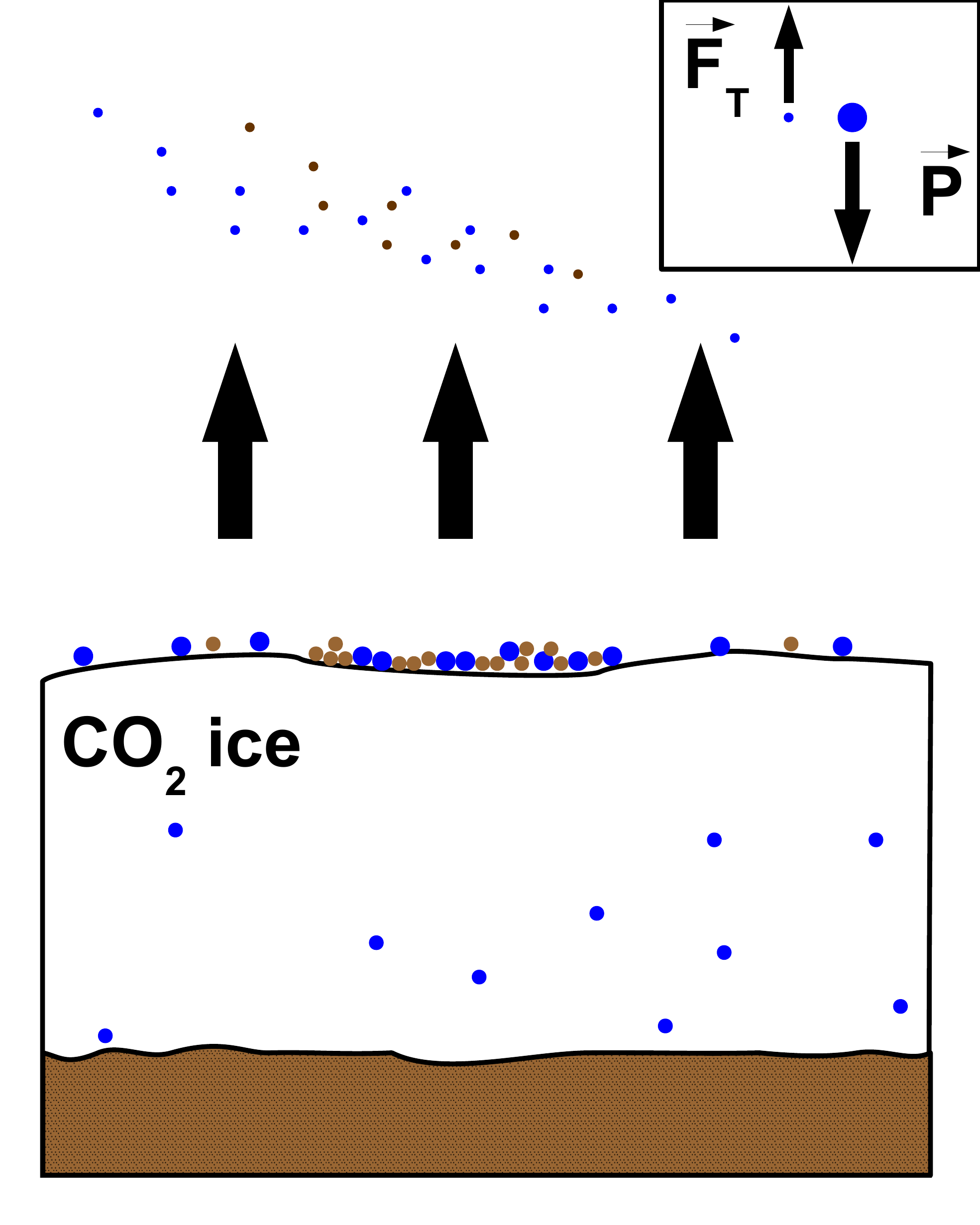}}
\par\end{centering}
\begin{centering}
\subfloat[]{\centering{}\includegraphics[width=0.33\textwidth]{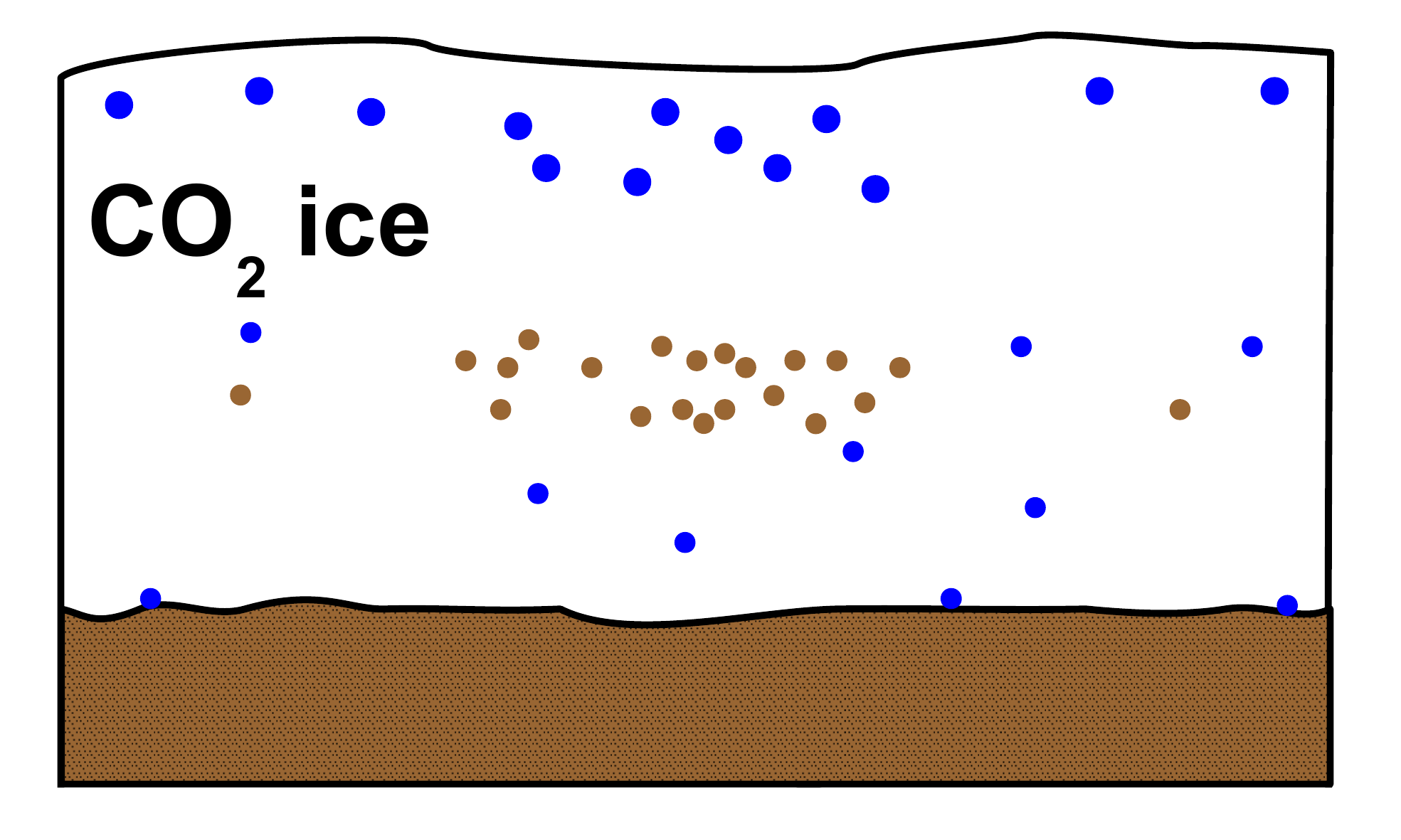}}\subfloat[]{\centering{}\includegraphics[width=0.33\textwidth]{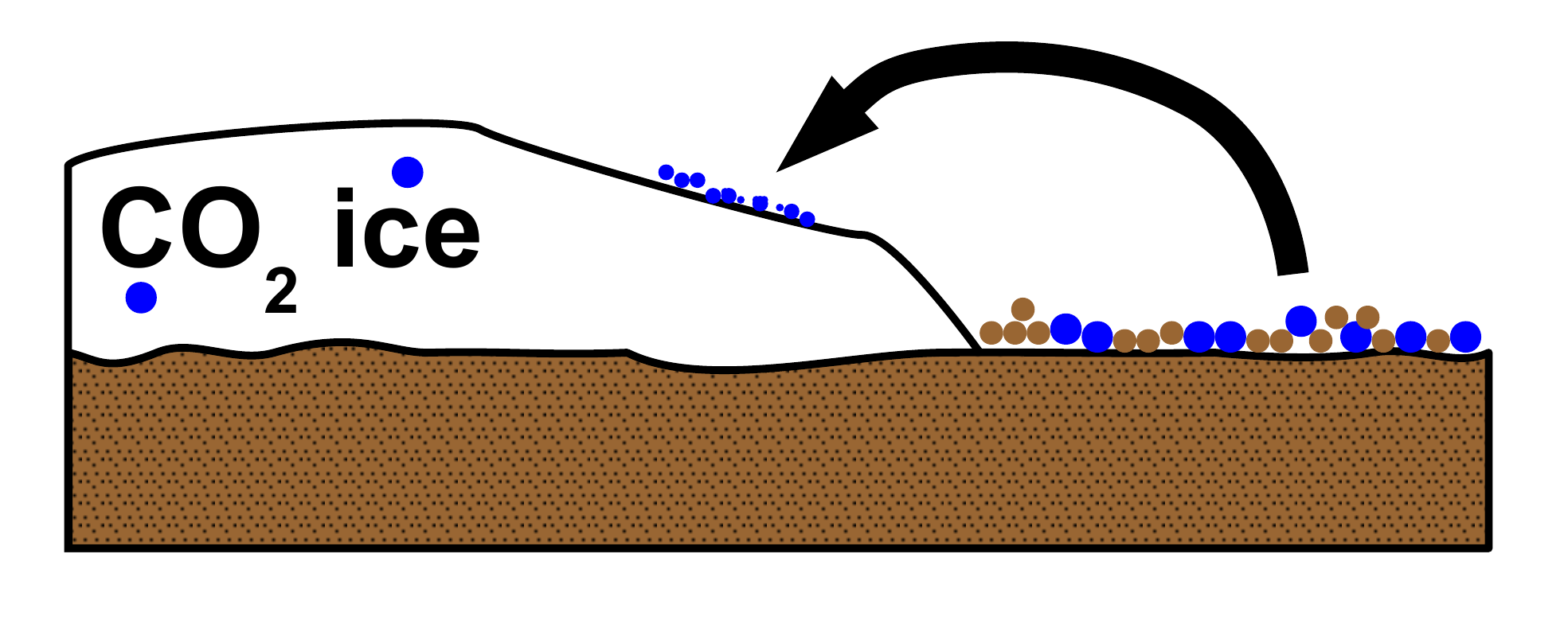}}\\
\par\end{centering}
\caption{\label{fig:Global_sketch}Proposed model for the evolution of seasonal
ice deposits during spring. (a) At the beginning of the spring the
ice layer becomes compact in slab form ($L_{S}\sim181\text{\textdegree}$),
In the meantime, it is cleaned of the impurities of dust and ice thanks
to solar irradiation (b) Basal ice sublimation starts, and the impurities
that were trapped in the ice are accumulated at the bottom of the
layer. This is the first stage of the cold geyser model. (c) The pressure
built inside pockets under the layer reaches the ice breaking threshold.
The ice breaks and gas escapes brutally, carrying impurities with
it. (d) Small grains are lifted and put into suspension into the atmosphere,
while large one fall back to the ground. Small water ice grains that
are put into suspension will eventually sublimate and be integrated
to the global water circulation. Dust is deposited in a superficial
layer. (e) In the place of a cold jet, the deposited layer of dust
and water ice is sinking. The dust sink very rapidly, and the water
ice is integrated to the layer due to the condensation of CO$_{2}$
at night at the top of the layer, and the sublimation during the day
of de CO$_{2}$ ice at the bottom of the layer. (f) If a patch of
terrain is completely defrosted, then the ground is no more thermally
buffered to the equilibrium temperature of CO$_{2}$ ice. The temperature
rises and the water ice sublimates. If CO$_{2}$ ice is still present
in the area, it will act as a cold trap, and the just sublimated water
will condense on the closest patch of CO$_{2}$ ice. }
\end{figure}

We propose a sketch (see fig. \ref{fig:Global_sketch}) to summarize
the evolution of the seasonal cap in Richardson crater in dark spots
and off dark spots.

\paragraph*{Mechanism for water escape}

We propose a mechanism of escape of water ice, which would be incorporated
into the atmosphere, from $L_{S}=190\text{\textdegree}$, in a way
consistent with the jet mechanism proposed by \citealp{Piqueux2003}
and \citealp{Kieffer2006,kieffer2007cold}.

If we consider that the sublimation flux can raise grains of dust
of a supposed density of 2200 $\mbox{kg}.\mbox{m}^{-3}$ up to a radius
of $3.5\mbox{\ensuremath{\mu}m}$ \citep{Kieffer2000,kieffer2007cold},
it is clear that water ice grains with a density of about 920 $\mbox{kg.m}^{-3}$
may be suspended to larger radii up to 10 microns. Step by step, all
the grains smaller than this radius will be eliminated (see Figure
\ref{fig:Global_sketch} d and e). Before the abrupt decrease $L_{s}=240\text{\textdegree}$,
the grain size seems to stabilize around a value between $10\,\mbox{\ensuremath{\mu}m}$
and $30\,\mbox{\ensuremath{\mu}m}$ (see fig. \ref{fig:WaterIceGrainsize}).
This value is consistent with the grain sizes that can escape.

\begin{figure}
\centering{}\includegraphics[width=1\textwidth]{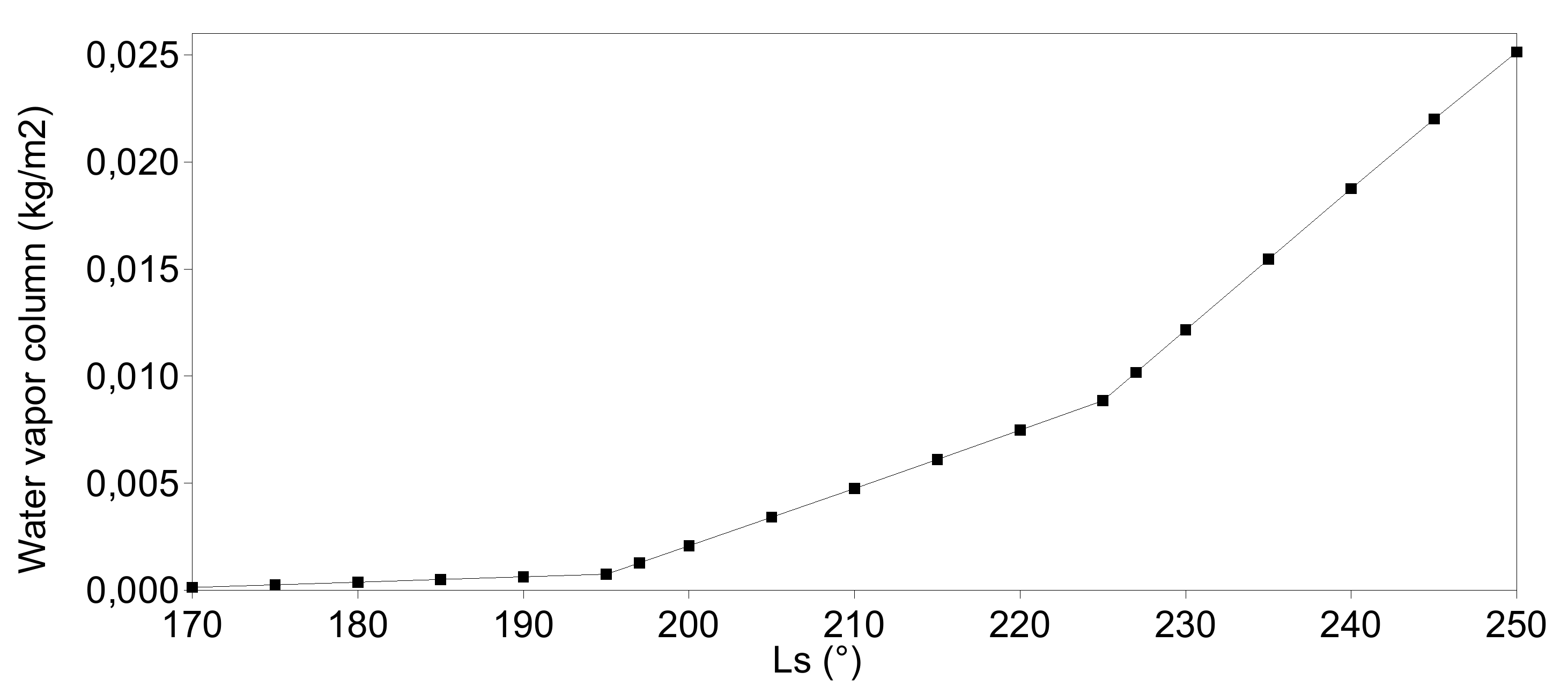}\caption{\label{fig:Contenu-en-vapeur}Atmospheric water vapor content in the
studied area, predicted by a GCM (\citealp{millour2014latest} ©LMD/OU/IAA/ESA/CNES). }
\end{figure}

OMEGA observations suggest an enrichment of the atmosphere at these
latitudes from $L_{S}=250\text{\textdegree}$ \citep{Melchiorri2009},
and are not incompatible with an enrichment earlier in the season.
It would be interesting to study these observations of atmospheric
water vapor in more detail. 

The water escape noted in section \ref{subsec:Impurities} is also
consistent with an enrichment of atmospheric vapor during the spring
predicted by climatic models \citep{Forget2006,millour2014latest},
as we can see it in figure \ref{fig:Contenu-en-vapeur}. This enrichment
is becoming more and more important as the season advances. This increase
in the quantity of water vapor predicted in the atmosphere by the
GCMs can be linked, on one hand, to the increase in atmospheric pressure
and, on the other hand, to forcing. 

\paragraph{Rolling layer mechanism}

According to MCD \citep{Forget99,Forget2006,millour2014latest}, at
the latitude of Richardson crater, $1.5\,\text{kg}$ of CO$_{2}$
ice recondenses each night per square meter at the beginning of the
spring. We can assume that this ice condenses mostly at the top of
the layer \citep{Pilorget2011} (because of the opacity of the CO$_{2}$
to thermal radiations, and because condensation is exothermic). This
means that from Ls=170\textdegree{} to Ls=250\textdegree , from the
beginning tho the end of the sublimation of the seasonal deposit,
a total amount of $112\,\text{kg}$ of CO$_{2}$ ice recondenses per
square meter meaning up to $70\,\text{mm}$ of CO$_{2}$ ice recondenses
at the surface. To determine this amount, we simply consider that
the amount of recondensed CO$_{2}$ decreases linearly to 0 at LS=250\textdegree .
This represents more than 15\% of the initial thickness of the layer.
This lead us to assume that the large water ice grains that are ejected
during a cold jet event, are re-integrated to the slab layer (see
figure\ref{fig:Global_sketch}e) by the condensation of CO$_{2}$
above them at night. 

\section{Conclusions and perspectives}

The main results provided by our data analysis are: (i) CO$_{2}$
ice is most probably in a translucent state during the whole spring,
instead of a granular one, as previously thought. (ii) Retrieved seasonal
variations of translucent ice thickness is consistent with GCM prediction.
(iii) Only few ppmv of dust are present in the CO$_{2}$ ice slab
during the spring. (iv) Since the water ice content is 0.1\%v throughout
the spring season but the CO$_{2}$ ice is shrinking, we must conclude
that a water escape occurs. The escape of water into the atmosphere
seems to be in agreement with water vapor measurements. (v) The water
grain size increases during the spring from 1 to 50 microns and suddenly
falls down again.

Using all these new constraints, we propose a new sketch to explain
the microphysics of the seasonal south polar cap. We propose a refined
cryoventing mechanism in the presence of water ice: the water ice
accumulates at the bottom of the layer as a result of the slab sublimation
from the bottom. Then water ice grains are ejected during cryoventing
episodes. The larger grains fall back on the layer, while the smaller
ones are lifted by the general sublimation flux, and integrated to
the atmospheric circulation. In the last stage, the water from already
defrosted regions condenses onto CO$_{2}$ ice that acts as a cold-trap. 

The small grain size of water that are incorporated into the general
circulation, should most probably sublimate. Future work, using for
instance limb measurements of MCS data, should confirm this mechanism.
Also the effect of this mechanism to the general circulation should
be studied by numerical simulations. Other study using CRISM data
in other location of the south pole, or even the north pole would
be interesting in order to determine if the proposed sketch is general
on Mars or particular to the Richardson crater. In this perspective,
regional scale analysis using OMEGA and the large footprint mode of
CRISM should also be conducted. 

\subsubsection*{Acknowledgement}

We acknowledge support from the ``Institut National des Sciences
de l'Univers'' (INSU), the \textquotedbl{}Centre National de la Recherche
Scientifique\textquotedbl{} (CNRS) and \textquotedbl{}Centre National
d'Etude Spatiale\textquotedbl{} (CNES) and through the \textquotedbl{}Programme
National de Plan\'etologie\textquotedbl{} and MEX/OMEGA Program.
The project « Multiplaneto » has been granted in 2015 and 2016 by
D\'efi Imag\textquoteright In/CNRS.

\bibliographystyle{elsarticle-harv}
\addcontentsline{toc}{section}{\refname}

\end{document}